\documentclass[epj,nopacs]{svjour}
\usepackage{pifont}
\usepackage{epsfig,array}
\usepackage{multirow,color}
\usepackage{epsf,graphicx}
\usepackage{times,latexsym,euscript,amstext,amssymb,amsbsy,amsmath,amsfonts}
\usepackage{mdwlist}\usepackage{epsfig}
\usepackage{slashbox,multirow}
\RequirePackage{lineno}
\newcommand{\be}{\begin{eqnarray}}
\newcommand{\ee}{\end{eqnarray}}
\newcommand{\bc}{\begin{center}}
\newcommand{\ec}{\end{center}}
\newcommand{\bea}{\begin{eqnarray}}
\newcommand{\eea}{\end{eqnarray}}

\newcommand{\beq}{\begin{equation}}
\newcommand{\eeq}{\end{equation}}

\begin{document}

% To get line numbers using lineo package
%\linenumbers

\title{\boldmath Energy-independent PWA of the reaction $\gamma p\to  K^+
\Lambda$ }
\titlerunning{Energy-independent PWA of the reaction
$\gamma p\to  K^+\Lambda$,}

\author{A.V.~Anisovich$\,^{1,2}$, R.~Beck$\,^1$, V. Burkert$\,^3$, E.~Klempt$\,^1$,
M.E.~McCracken$\,^{4,5}$, V.A.~Nikonov$\,^{1,2}$,
A.V.~Sarantsev$\,^{1,2}$, R.A.~Schumacher$\,^5$, U.~Thoma$\,^1$ }
\authorrunning{A.V.~Anisovich \it et al.}

\institute{$^1\,$Helmholtz-Institut f\"ur Strahlen- und Kernphysik, Universit\"at Bonn, Germany\\ 
$^2\,$Petersburg Nuclear Physics Institute, Gatchina, Russia\\ 
$^3\,$Jefferson Lab, 12000 Jefferson Avenue, Newport News, Virginia, USA \\ 
$^4\,$Washington \& Jefferson College, Washington, Pennsylvania, USA \\
$^5\,$Carnegie Mellon University, 5000 Forbes Ave., Pittsburgh, Pennsylvania 15213, USA }

\date{Received: \today / Revised version:}

\abstract
{Using all recent data on the differential cross sections and spin observables for the reaction $\gamma p \to K^+ \Lambda$, an energy-independent partial-wave analysis is performed.
The analysis requires multipoles up to $L = 2$; there is no evidence that the fit requires multipoles with $L=3$. At present the available data allow us to  extract the dominant multipoles only. These are compatible with the multipoles obtained in the energy-dependent fit.
This result supports the reliability of the Bonn-Gatchina energy-dependent results.}
\maketitle

\section{Introduction}
The spectrum of excited nucleons ($N^*$) reflects the
structure of quantum chromodynamics in the non-perturbative regime.
Full elucidation of the properties of the $N^*$ states is an important
long-term goal for the hadron physics community.  The
majority of experimental information has historically come from
nucleon and pion-induced elastic and inelastic reaction channels.
Partial-wave analyses of these data have established a rich spectrum
of $N^*$ states, so that the Particle Data Group \cite{PDG:2012}
lists, for example, well-established ({\em i.e.}, ``existence is
certain'' rating) nucleon resonances with spins up to $9/2$.  In recent
years, new experiments have produced high-precision measurements of
photo-production of several hadronic final states due to the
availability of high quality photon and electron beams at facilities
including CLAS/Jefferson Lab, ELSA/Bonn, MAMI/ Mainz, LEPS/SPring-8,
and GRAAL/Grenoble.  These have had a significant impact on our
understanding of $N^*$ properties.  The present paper dwells on the
production of a strangeness-containing final state. The reaction
\begin{equation}
\label{eq:1}\gamma p \rightarrow K^+\Lambda
\end{equation}
is sensitive to $N^*$ states with masses in the range from 1.6 to
2.4~GeV, a two-body final state in a domain where pionic reactions are
dominated by more complicated multi-pion final states.  This makes it
attractive to study, since the simple kinematics gives straightforward
access to non-strange excitations in a mass range that is otherwise
not well understood.

Theoretical work using a relativized constituent quark model
made predictions about the spectrum of $N^*$ and $\Delta$
excitations \cite{Capstick:1993piN,Capstick:1998D}, as
well as their couplings to various initial and final states including
hyperonic~\cite{Capstick:1998KY} states.
Many of the observed $N^*$
resonances, as well as many ``missing'' resonances that have not been
observed coupling to $\pi N$ were tabulated. One goal of $N^*$ spectroscopy
programs is thus to search for such missing states in channels other
than $\pi N$.

All pseudo-scalar meson photoproduction reactions are
characterized by eight complex amplitudes; parity invariance of the
strong interaction reduces this number to four independent
amplitudes. Full characterization of the reaction at a given kinematic
relies upon measurement of $d\sigma/d\Omega$ and at least seven of the
fifteen single- and double-polarization observables~\cite{Chiang:1996em}.
However, for the sake of redundancy and the reduction of experimental
ambiguities, as broad a range of observables as feasible must be
analyzed for a full decomposition of a reaction at the amplitude
level. In most theoretical or phenomenological studies, these amplitudes are constructed for each bin in energy and 
angle. In a next step, the set of amplitudes at a given energy can 
be expanded into multipoles which contain the information on the 
underlying physical processes. 

Less demanding is to use only a finite number of multipoles in a 
truncated partial-wave expansion of the photoproduction amplitude. 
For small numbers of contributing  partial waves, five observables 
can already be sufficient to determine the amplitudes 
\cite{Omelaenko:1981,Wunderlich:2013iga}. For example, seven observables 
have been measured for $\gamma p \to \pi^0p$, the differential cross 
section $d\sigma/d\Omega$, the beam asymmetry $\Sigma$, target 
asymmetry $T$, the recoil polarization $P$, and different correlations 
between photon and target polarization yielding $G$, $E$, and $H$. 
The seven data sets span a common mass range from 1.462 to 1.662\,GeV 
in which no contributions with orbital angular momenta $L\ge 3$ are 
expected. A truncated partial-wave analysis returned multipoles with 
$L=0,1$ and $2$ \cite{Hartmann:2014}. These multipoles lead to the
excitation of nucleon and $\Delta$ resonances. In the 1500\,MeV region, however, 
the impact of $\Delta$ resonances is small, and that is why 
the $N(1520)3/2^{-}$ helicity coupling could be determined in 
\cite{Hartmann:2014}.  

The $\gamma p \rightarrow K^+\Lambda$ reaction profits from the fact that - due to isospin 
conservation - only isospin-1/2 intermediate states contribute and thus 
all $\Delta$ excitations are excluded. Furthermore, the weak decay of 
the $\Lambda$ to $\pi N$ allows for determination of its recoil 
polarization.  Use of polarized photon beams gives access to other 
polarization observables, notably the beam asymmetry ($\Sigma$), and beam-recoil
double-polarization observables for both circular ($C_{x}$, $C_{z}$)
and linear ($O_{x}$, $O_{z}$) photon polarizations. Recent and
forthcoming measurements from experiments with polarized nucleon
targets will give access to the remaining set of polarization
observables.  For these reasons, the $\gamma p \rightarrow K^+\Lambda$
reaction is presently the best candidate for full amplitude-level
characterization.

So far, all partial-wave analyses (PWA) of the reaction $\gamma p \rightarrow K^+\Lambda$ used 
energy-dependent representations of the contributing $N^*$ states in 
the reaction. We shortly review these analyses in Sec.~\ref{sec:analyses}. 
Energy-dependent analyses benefit from the analytic structure of the 
amplitudes. However, the dominant partial waves over a range of energies 
could tend to mask the contributions of weaker partial waves nearby due 
to the force of statistics. A crucial test of the validity of this 
approach is to compare it to the results of an energy-independent 
analysis. The two approaches should agree within their respective 
limitations. Here we present application of the Bonn-Gatchina (BnGa) model to 
the $\gamma p \rightarrow K^+\Lambda$ reaction data in eleven 
independent energy bins from threshold up to $\sqrt{s} = 1918$~MeV, 
as itemized in Sec.~\ref{sec:data}.  The formalism is reviewed in
Sec.~\ref{sec:formalism}.  Results are discussed in
Sec.~\ref{sec:results}, where we compare the fits with the latest 
energy-dependent BnGa multi-channel PWA. We
also compare the energy-independent fits with multipolarity $L = 0, 1$
with those including $L=2$ to assess the need for higher $J^P$
intermediate states in the $\gamma p \rightarrow K^+\Lambda$ reaction.
We show how using the energy-independent solution can be used to check
the stability of earlier energy-dependent fit results.
Sec.~\ref{sec:conclusions} summarizes our findings.

%Study of the reaction
%\be
%\gamma p\to \Lambda K^+ \label{eq:1}
%\ee
\section{Energy-dependent analyses}
\label{sec:analyses}

In light of the several attractive features of the $\gamma p \rightarrow  K^+\Lambda$ reaction, it has been the most suitable candidate for
partial-wave analysis (PWA); several analyses have been performed with
varied techniques and results.  Early analyses \cite{Mart:1999}
applied a single-channel tree-level resonant isobar model to SAPHIR
$d\sigma/d\Omega$ data \cite{Glander:2004} and found contributions
from known $N(1650)1/2^-$, $N(1710)1/2^+$, and $N(1720)3/2^+$ states,
as well as evidence for a previously unobserved $3/2^-$ state with a
mass of 1894~MeV.  Soon after, other work showed features of the
SAPHIR data that had been interpreted as resonant contributions could
be described in a Regge model \cite{Guidal:2003} to describe
$t$-channel exchange of strange mesons \cite{Janssen:2001}.  The group
at Ghent~\cite{Corthals:2006} used the Regge-plus-resonance approach to
analyze forward-angle $d\sigma/d\Omega$ and $P$ data and found
evidence for $N(1650)1/2^-$, $N(1710)1/2^+$, and $N(1720)3/2^+$ states
near threshold and $J^P = 3/2^+$ and $1/2^+$ states with masses near
1.9 GeV.  Exploratory PWA studies at $\sqrt{s} > 2.0$~GeV indicate
resonance structure near 2.1 GeV, for example in~\cite{Schumacher:2010qx}, but in the present study we only work
with multipoles below about 1.92 GeV.

%A dynamical
%coupled-channel analysis with data on the $\gamma p \rightarrow p
%\eta$ reaction \cite{Julia-Diaz:2005} found the need for a $J^P =
%1/2^-$ state at $\sqrt{s}\approx 1.9$~GeV.

Interpretation of $K^+\Lambda$ production mechanism was complicated
when $d\sigma/d\Omega$ and recoil polarization data published by the
CLAS Collaboration~\cite{Bradford:2005pt,McNabb:2003nf} showed
significant differences when compared to the SAPHIR data.  The
implications were studied and discussed by Mart {\it et
al.}~\cite{Mart:2006dk} and others in both single-channel effective
Lagrangian models and multipole analyses.  

The Bonn-Gatchina 
Group produced several PWA studies over the years of the $\gamma p
\rightarrow K^+\Lambda$ reaction.  A 2005 publication Sarantsev {\it
et al.}~\cite{Sarantsev:2005} demonstrated that partial-wave analysis
of $d\sigma/d\Omega$, $\Sigma$ and $P$ data, when coupled with data
from photoproduction of $K\Sigma$, $\pi N$, and $\eta N$, necessitates
a $J^P = 1/2^+$ state with mass of approximately 1840~MeV. It also suggested
the existence of four $3/2^-$ states between 1520 and 2170~MeV, but
produces no evidence for $1/2^-$ states of mass above 1650~MeV.  They
noted that the discrepancy between the then available SAPHIR and CLAS
$d\sigma/d\Omega$ results could lead to ambiguities in fitting.  With the
publication~\cite{Anisovich:2007cxcz},  Anisovich {\it et al.}
incorporated the large spin-transfer probability observables $C_x$ and
$C_z$ measured with circularly polarized photons at CLAS
\cite{Bradford:2006ba} into an analysis coupling several observables
from $\pi$, $\eta$, and $K$ photoproduction reactions.  This analysis
showed that all observables could be reproduced with the further
addition of only one state, a $3/2^+$ resonance with mass of
approximately 1900~MeV.  The 2010 publication of higher-statistics and
independent results from CLAS
\cite{McCracken:2009ra}, the discrepancy in the cross section for $\gamma p \rightarrow K^+
\Lambda$ seems to have been resolved. Reduced
ambiguity in PWA of the channel was thus expected.  

The most recent
BnGa analyses have applied coupled-channel PWA to a large set of
observables for many reactions \cite{Anisovich:2011fc}, and extracted
transition amplitudes for pion- and photon-induced production of
$\eta$ and $K$ mesons \cite{Anisovich:2012ct}.  These analyses show
that adequate description of the data is possible with two separate
sets of resonances, distinguished by the presence of either one or two
$7/2^+$ states.  A subsequent multi-channel analysis
\cite{Anisovich:2013vpa} focusing on information from $K\Sigma^0$
production, provided an updated set of resonance contributions
(referred to here as BnGa2013), but concluded that further polarization
information for $\Sigma^0$ production is needed to unambiguously
determine production amplitudes for these reactions.

\section{\boldmath Data on $\gamma p\to K^+\Lambda $}
\label{sec:data}

\begin{figure*}[pt]
\begin{center}
\begin{tabular}{ccc}
\hspace{-2mm}\includegraphics[width=0.42\textwidth]{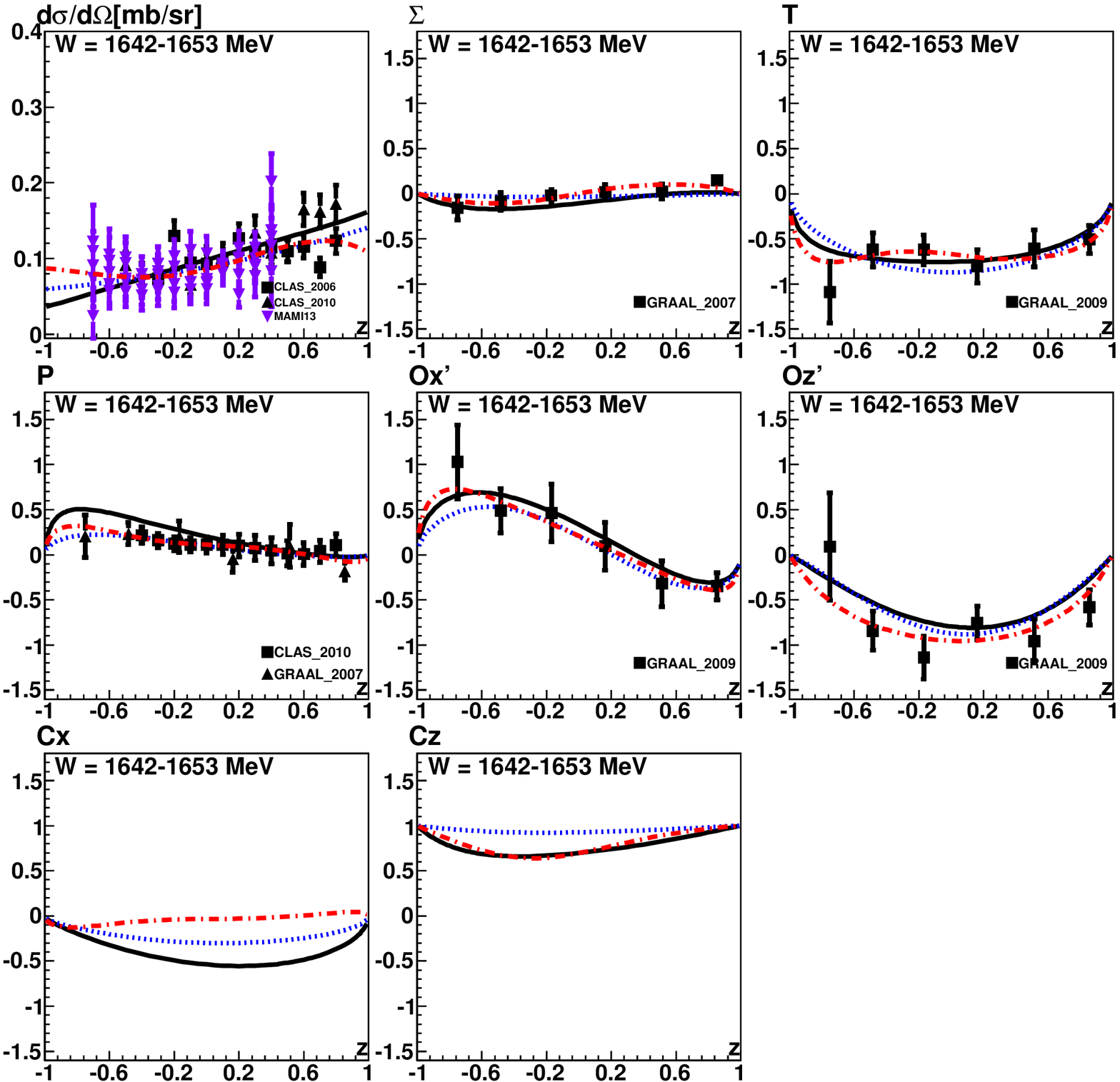}&
\includegraphics[width=0.42\textwidth]{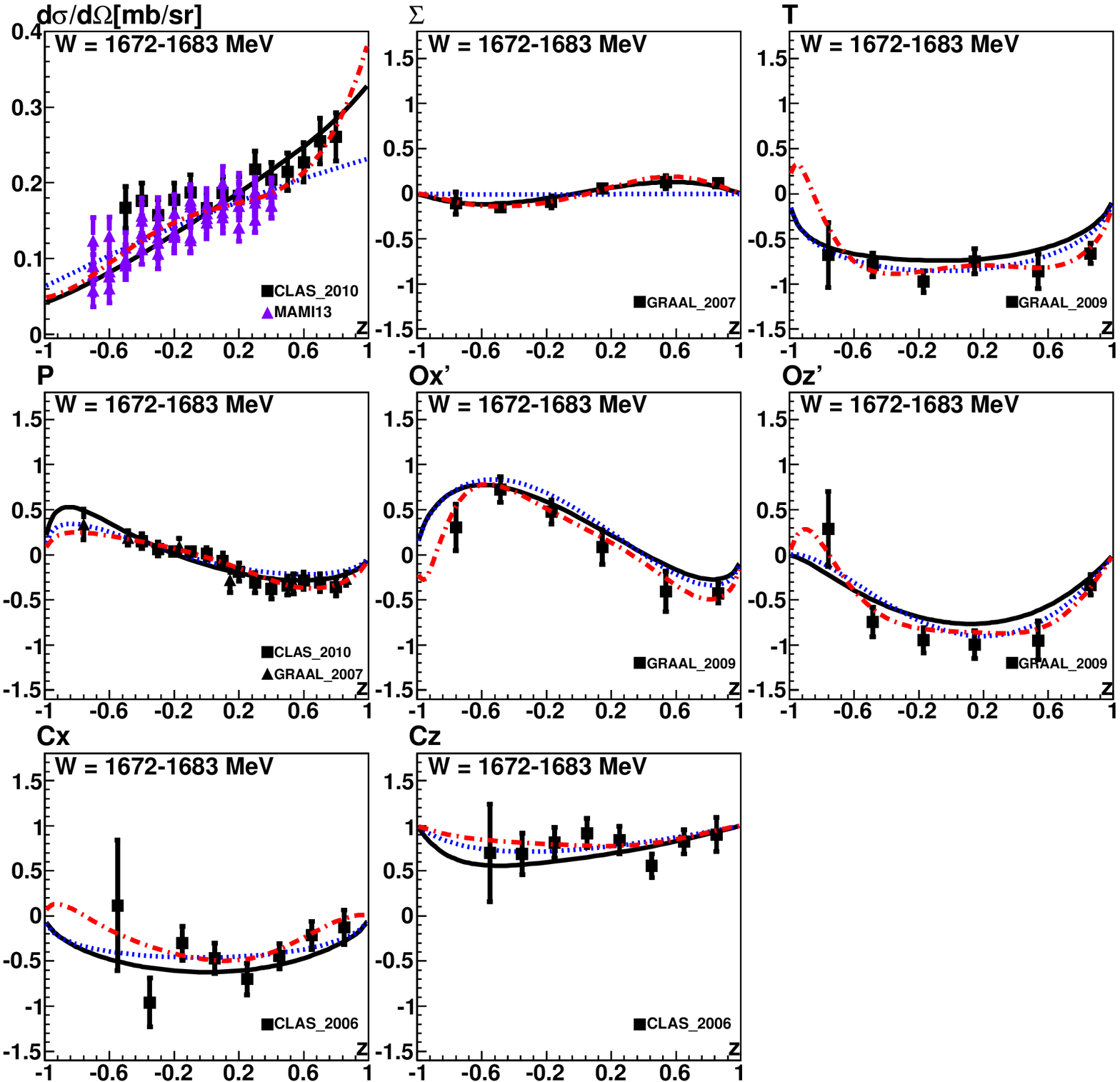}\\[2ex]
\hspace{-2mm}\includegraphics[width=0.42\textwidth]{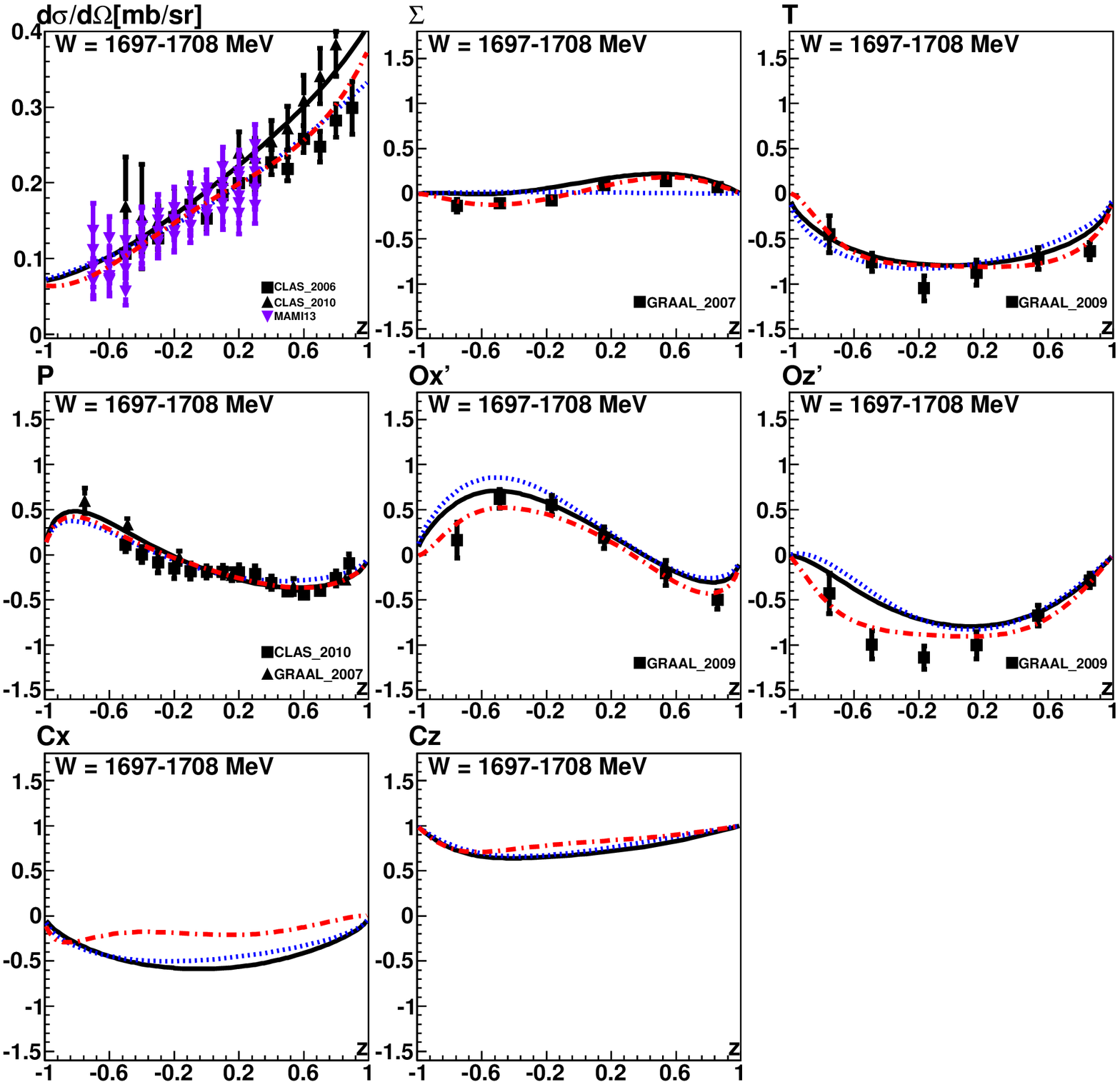}&
\includegraphics[width=0.42\textwidth]{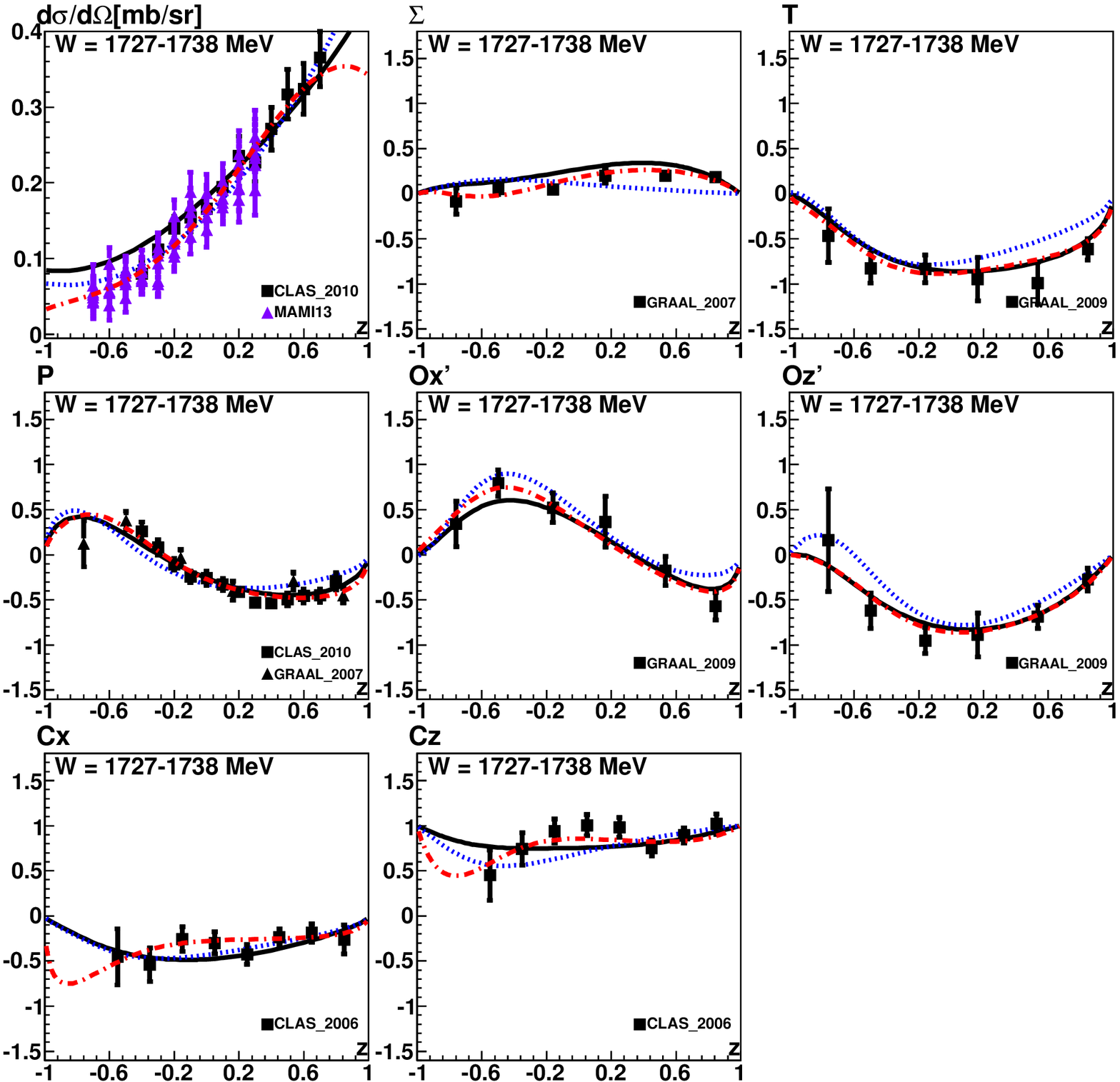}\\[2ex]
\hspace{-2mm}\includegraphics[width=0.42\textwidth]{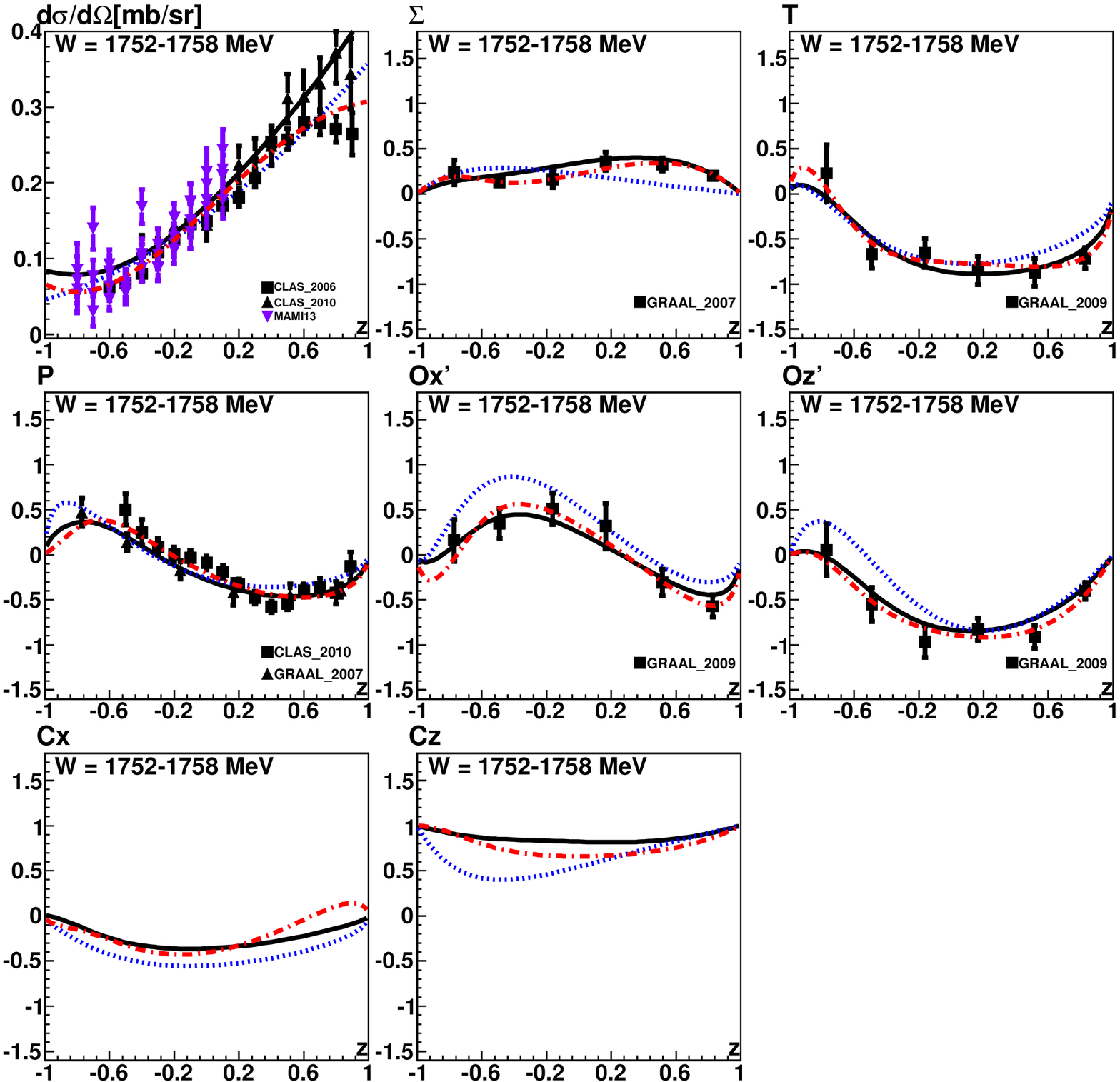}&
\includegraphics[width=0.42\textwidth]{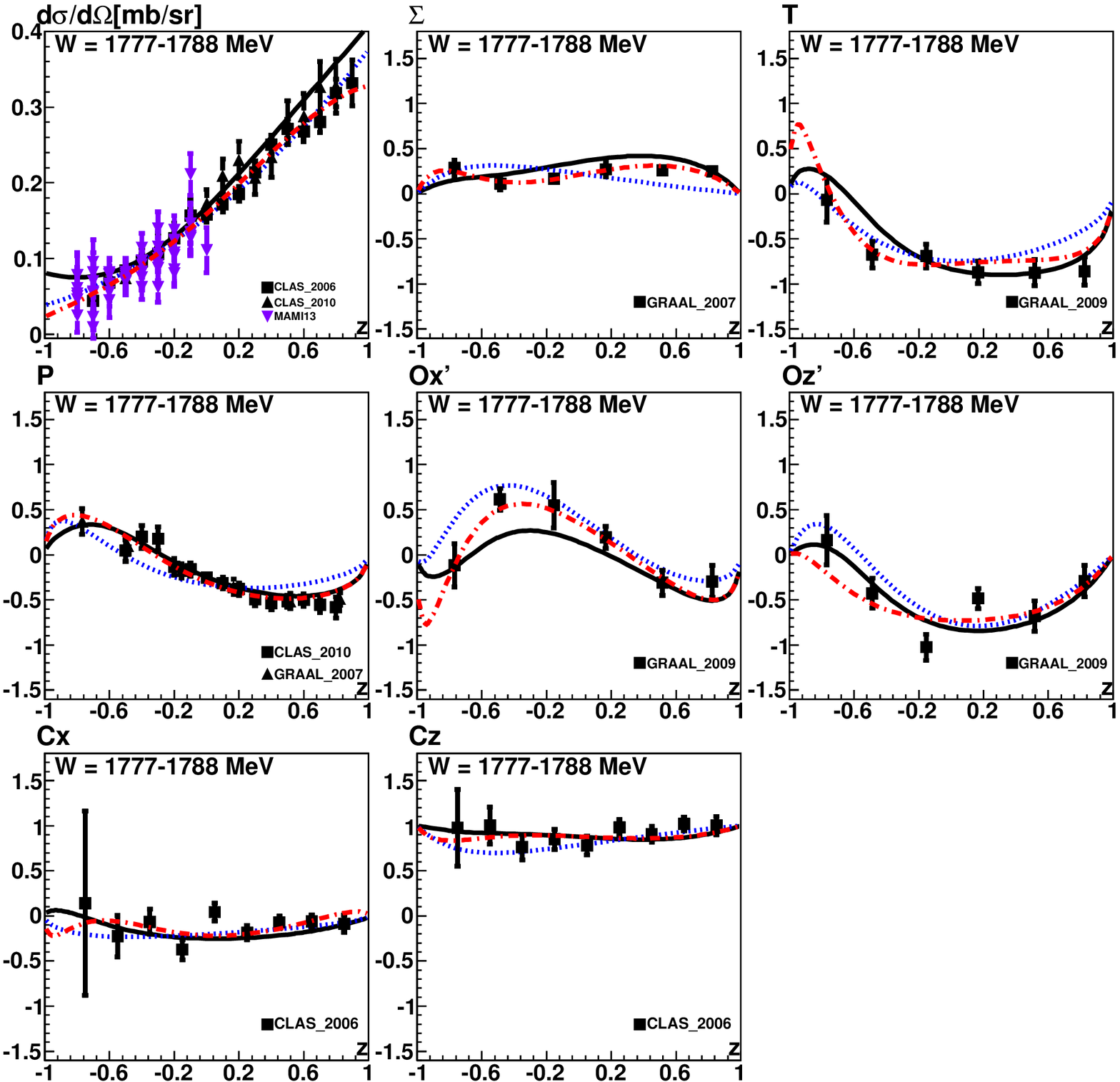}
\end{tabular}
\end{center}
\caption{
\label{Observables}
Data and fit to data on $\gamma p\to  K^+\Lambda$. Black line is the
energy-dependent solution BnGa2013, the dashed (blue) line is the
truncated PWA with $L=0,1$, and dot-dashed (red) line the truncated
PWA with $L=0,1,2$. }
\end{figure*}

In the region from threshold up to $\sqrt{s}=1918$~MeV the following data are used: differential cross section $d\sigma/d\Omega$ from \cite{Bradford:2005pt} and \cite{McCracken:2009ra}, recoil polarization $P$ from \cite{McCracken:2009ra} and \cite{Lleres:2007tx}, 
$\Sigma$ from \cite{Lleres:2007tx}, $T$, $O_{x'}$ and $O_{z'}$ from \cite{Lleres:2008em}, and the spin transfer coefficients $C_x$ and $C_z$ from \cite{Bradford:2006ba}. In the low energy bin the recent data on $d\sigma/d\Omega$ from CB@MAMI \cite{Jude:2013jzs} are also used. 
The fitting region was divided into eleven energy bins in $\sqrt{s}$ each containing data on at least one double polarization observable: $1642-1653, 1672-1683, 1697-1708, 1727-1738,
1752-1758, 1777-1788, 1807-1818$, $1832 -1843, 1857-1868, 1882-1893, 1900-1918$ (in MeV). The data are shown in Fig.~\ref{Observables} jointly with the BnGa2013 energy-dependent fit and two further fits described below. The data are organized in eleven blocks each containing the eight 
observables. 

\begin{figure*}[pt]
\begin{center}
\begin{tabular}{ccc}
\hspace{-2mm}\includegraphics[width=0.42\textwidth]{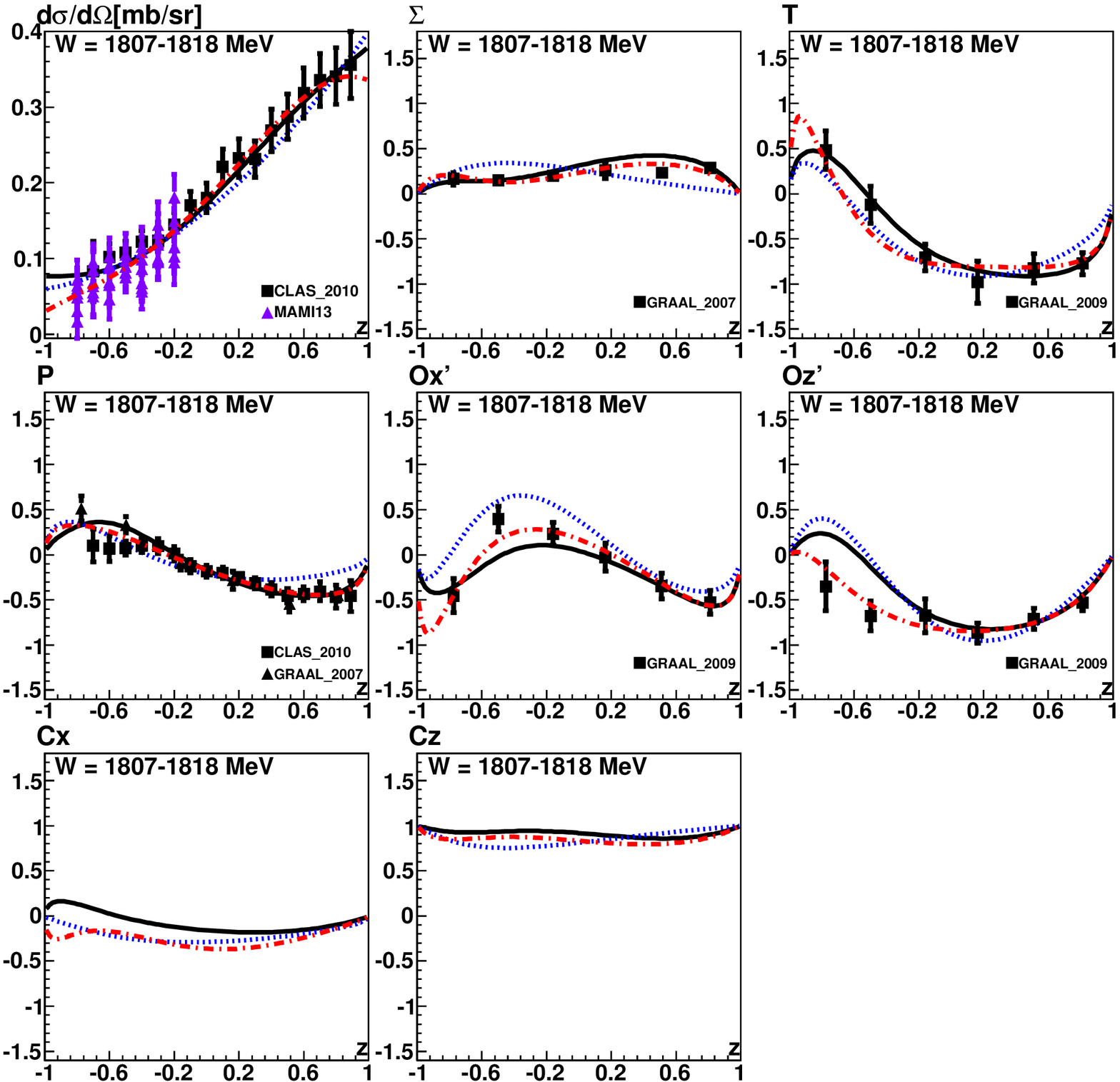}&
\includegraphics[width=0.42\textwidth]{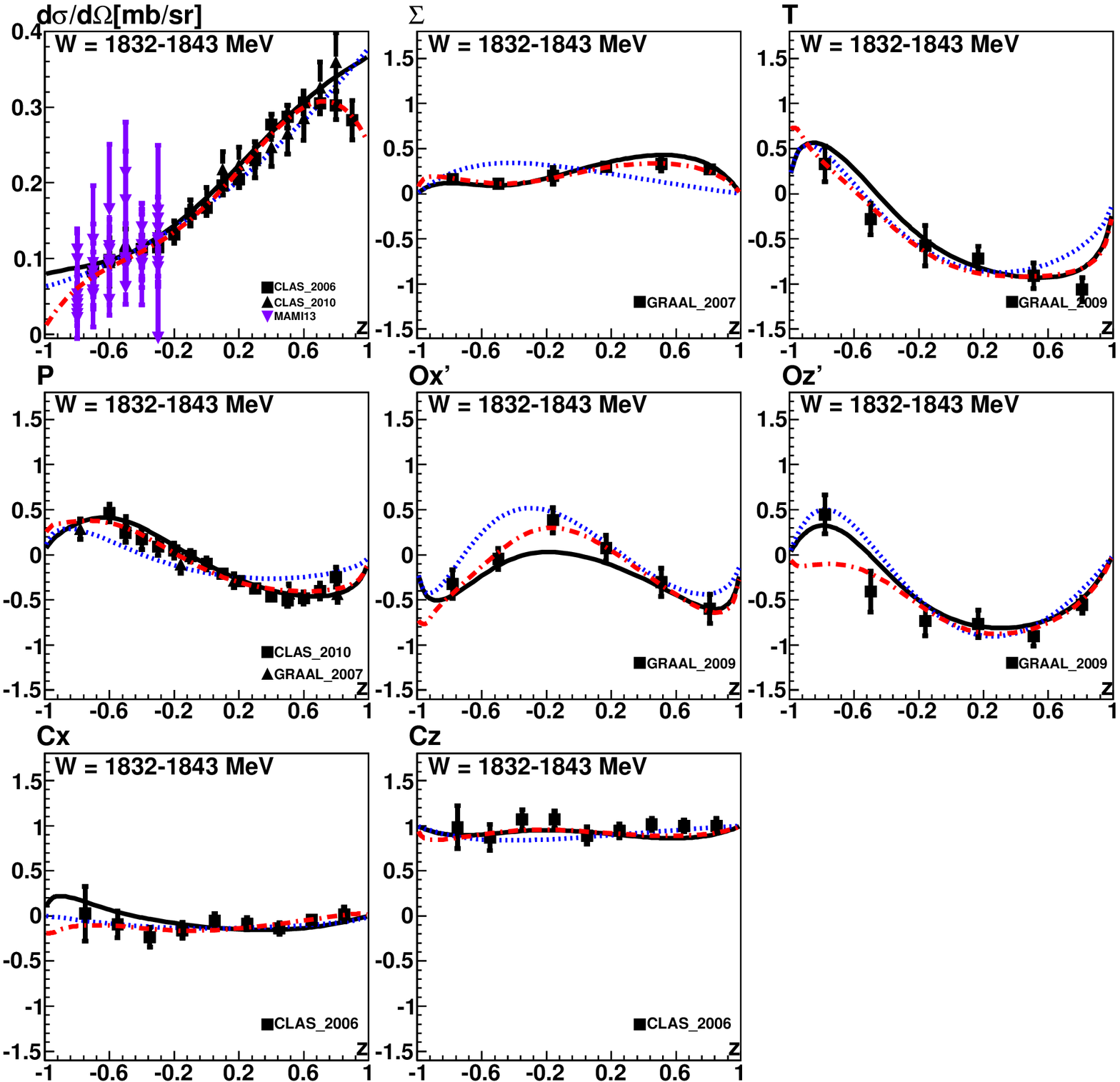}\\[2ex]
\hspace{-2mm}\includegraphics[width=0.42\textwidth]{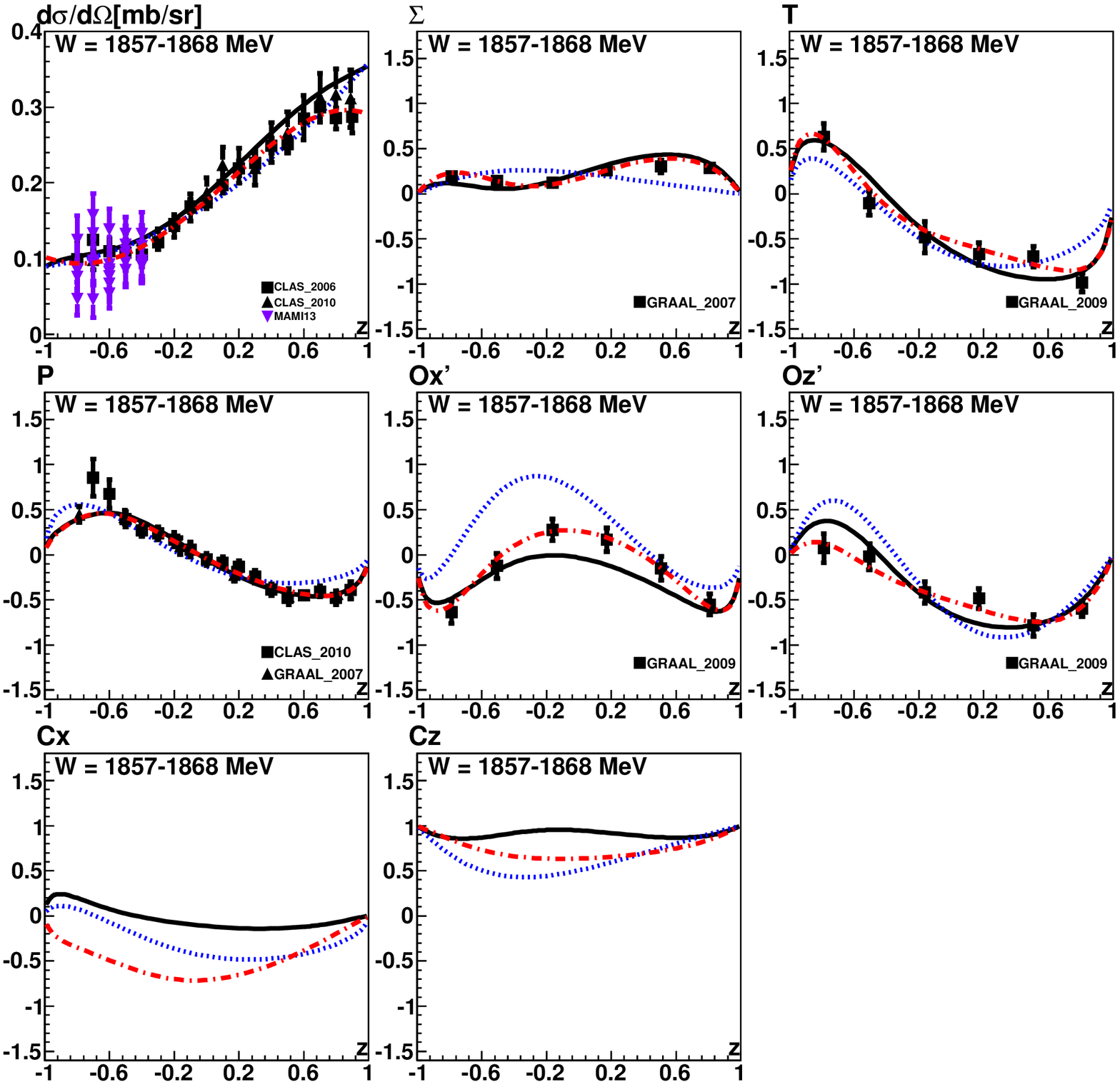}&
\includegraphics[width=0.42\textwidth]{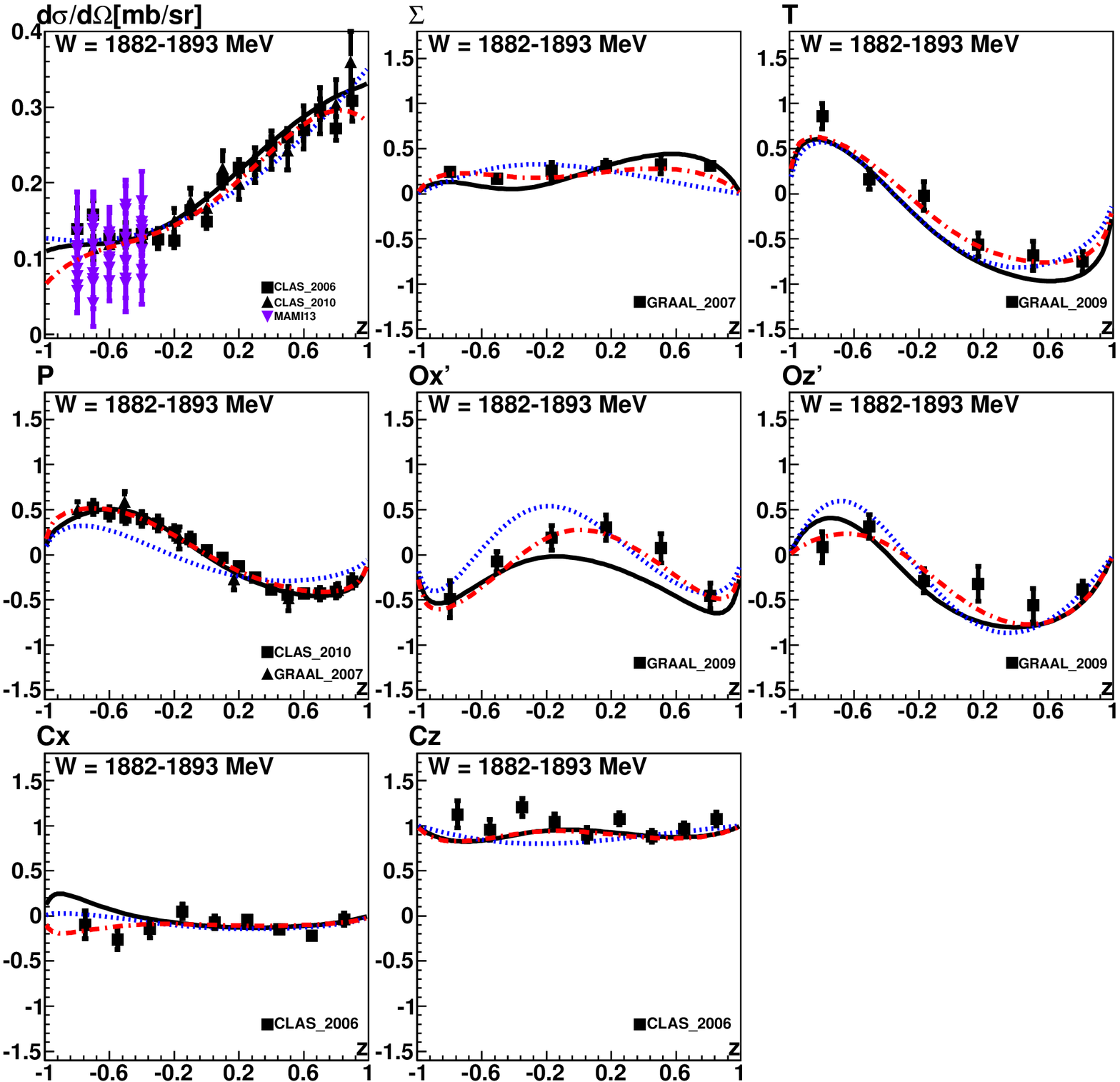}\\[2ex]
\hspace{-2mm}\includegraphics[width=0.42\textwidth]{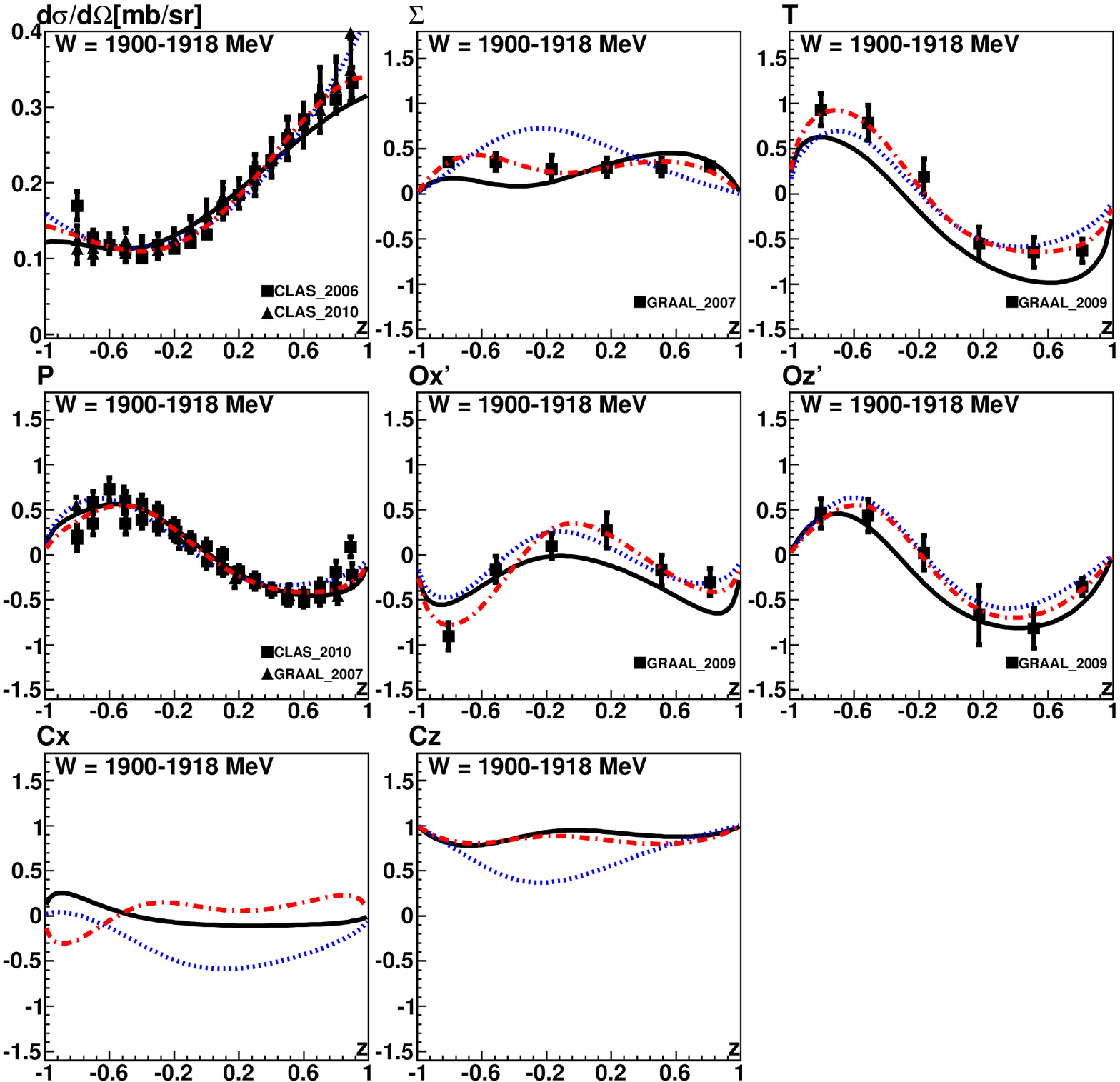} 
\end{tabular}
\end{center}
{\bf Fig. 1 continued.}
\end{figure*}
\section{Formalism}
\label{sec:formalism}

The amplitude for photoproduction of a single pseudoscalar meson is
well known and can be found in the literature (see for example
\cite{Drechsel:1998hk} and references therein). Here, we consider
the case of a $K^+$ meson recoiling against a $\Lambda$ hyperon. The
general structure of the amplitude can be written in the form
$$A=\omega^*J_\mu\varepsilon_\mu \omega' \;,$$
where $\omega'$ and $\omega$ are spinors representing the baryon in the 
initial and final state, $J_\mu$ is the electromagnetic current of the electron, and $\varepsilon_\mu$ characterizes the polarization of the photon.
The full amplitude can be expanded into four invariant (CGLN) amplitudes $\mathcal F_i$~~\cite{Chiang:1996em}
\be
\label{mult_1}
&& J_\mu = \\&& i {\mathcal F_1}
 \sigma_\mu +{\mathcal F_2} (\vec \sigma \vec q)
\frac{\varepsilon_{\mu i j} \sigma_i k_j}{|\vec k| |\vec q|} +i {\mathcal F_3} \frac{(\vec \sigma
\vec k)}{|\vec k| |\vec q|} q_\mu +i {\mathcal F_4} \frac{(\vec \sigma \vec q)}{\vec q^2} q_\mu
\;.\nonumber
\ee
where $\vec q$ is the momentum of the nucleon in the $ K^+\Lambda$
channel, $\vec k$ is the momentum of the nucleon in the $\gamma N$
channel calculated in  the center-of-mass system of the reaction,
and $\sigma_i$ are the Pauli matrices.

The functions ${\mathcal F_i}$ have the following angular dependence:
\be
&{\mathcal F_1}(W,z) &= \nonumber
 \sum^{\infty}_{L=0} [LM_{L+}+E_{L+}]
P^{\prime}_{L+1}(z) + \\&& [(L+1)M_{L-} + E_{L-}] P^{\prime}_{L-1}(z) \;, \nonumber \\& {\mathcal
F_2}(W,z)& = \sum^{\infty}_{L=1} [(L+1)M_{L+}+LM_{L-}] P^{\prime}_{L}(z)   \;,\nonumber  \\&
{\mathcal F_3}(W,z)& =  \sum^{\infty}_{L=1} [E_{L+}-M_{L+}] P^{\prime\prime}_{L+1}(z) + \\&&[E_{L-}
+ M_{L-}] P^{\prime\prime}_{L-1}(z)\;, \nonumber \\& {\mathcal
 F_4} (W,z) &= \sum^{\infty}_{L=2} [M_{L+}
- E_{L+} - M_{L-} -E_{L-}] P^{\prime\prime}_{L}(z). \nonumber
\label{mult_2}
\ee
Here, $L$ corresponds to the orbital angular momentum in the
$ K^+\Lambda$ system,  $W$ is the total energy, $P_L(z)$ are
Legendre polynomials with $z = (\vec k\vec q)/(|\vec  k||\vec q|)$,
and $E_{L\pm}$ and $M_{L\pm}$ are electric and magnetic multipoles
describing transitions to states with $J=L\pm 1/2$. There are no
contributions from $M_{0+}$, $E_{0-}$, and $E_{1-}$ for spin 1/2
resonances.

Differential cross section and polarization observables can be
expressed in terms of the ${\mathcal F_i}$ functions.  The relations
can be found, e.g., in \cite{Fasano:1992es}. For convenience, we
give the expressions for the observables used in the fit. The single
polarization observables $\Sigma$, $P$ and $T$ are given by
\be
  \Sigma\; I &=&
     -\frac{\sin^2(\theta)}{2}\\&&\hspace{-4mm} Re[{\mathcal F_3} {\mathcal F_3^*} + {\mathcal F_4} {\mathcal F_4^*}+
             2{\mathcal F_4} {\mathcal F_1^*} + 2{\mathcal F_3} {\mathcal F_2^*} +2 z {\mathcal F_4} {\mathcal
             F_3^*}]\,, 
             \nonumber\\
  P\;I &=&
         \sin(\theta) Im[(2 {\mathcal F_2^*} + {\mathcal F_3^*}+ z {\mathcal F_4^*}) {\mathcal F_1} + \\&&
          {\mathcal F_2^*}(z {\mathcal F_3} + {\mathcal F_4}) +
                          \sin^2(\theta) {\mathcal F_3^*} {\mathcal F_4}]\,, \nonumber\\
  T\;I &=&
           \sin(\theta)Im[{\mathcal F_1^*} {\mathcal F_3} - {\mathcal F_2^*} {\mathcal F_4} + \\&&
            z({\mathcal F_1^*} {\mathcal F_4} - {\mathcal F_2^*} {\mathcal F_3}) -
                          \sin^2(\theta) {\mathcal F_3^*} {\mathcal F_4}]\,, \nonumber
\ee
where
\be
    &&I = Re[{\mathcal F_1} {\mathcal F_1^*} + {\mathcal F_2} {\mathcal F_2^*} -2 z {\mathcal F_2} {\mathcal F_1^*} + \\&&\hspace{-4mm}
    \frac{\sin^2(\theta)}{2}
             ({\mathcal F_3} {\mathcal F_3^*} + {\mathcal F_4} {\mathcal F_4^*} + 2 {\mathcal F_4} {\mathcal F_1^*} +
              2 {\mathcal F_3} {\mathcal F_2^*} + 2 z {\mathcal F_4} {\mathcal F_3^*})] \nonumber .
\ee
Here the center of mass (c.m.) scattering angle is $\theta$.
The double polarization observables $O_{x'}$, $O_{z'}$, $C_{x}$ and
$C_{z}$ can be written as
\be
  O_{x'}\;I &=&\\
           &&\hspace{-4mm}\nonumber\sin(\theta) Im[{\mathcal F_2} {\mathcal F_3^*} - {\mathcal F_1} {\mathcal F_4^*}
                                         + z ({\mathcal F_2} {\mathcal F_4^*} - {\mathcal F_1} {\mathcal
                                         F_3^*})]\,,   \\
  O_{z'}\;I &=&
           -\sin^2(\theta) Im[{\mathcal F_1} {\mathcal F_3^*} + {\mathcal F_2} {\mathcal F_4^*}]\,,\\
 C_{x} &=& \sin(\theta) C_{z'} + \cos(\theta) C_{x'}\,,\\
 C_{z} &=& \cos(\theta) C_{z'} - \sin(\theta) C_{x'}\,,
\ee
where
\be
  C_{x'}\;I &=& \sin(\theta) Re[{\mathcal F_2} {\mathcal F_2^*} - {\mathcal F_1} {\mathcal F_1^*} +
   {\mathcal F_2} {\mathcal F_3^*} - {\mathcal F_1} {\mathcal F_4^*}
                             + \nonumber \\\,,&&
                             z ({\mathcal F_2} {\mathcal F_4^*} - {\mathcal F_1} {\mathcal F_3^*})]\,, \\
  C_{z'}\;I & =&  Re[-2 {\mathcal F_1} {\mathcal F_2^*} + z ({\mathcal F_1} {\mathcal F_1^*} +
   {\mathcal F_2} {\mathcal F_2^*})- \nonumber \\&&
                              \sin^2(\theta) ({\mathcal F_1} {\mathcal F_3^*} + {\mathcal F_2} {\mathcal
                              F_4^*})]\,.
\ee
Let us remind the reader that the $z$ axis defines the direction of the incoming
particles in the c.m. system, while the $z'$ axis defines the direction of the
outgoing particles (see \cite{Fasano:1992es}). Finally the
differential cross section is equal to:
\be
\frac{d\sigma}{d\Omega}\;=\;\frac{k}{q}\;I\,,
\ee
where $q$ and $k$ are the 
moduli of the initial and final c.m.~momenta, respectively.

%===========================================================================================================

\section{Energy-independent truncated PWA}
\label{sec:results}

The energy-independent (or single energy) PWA uses the full database
of the Bonn-Gatchina partial-wave analysis \cite{Anisovich:2011fc}. A
reasonable description of all data is achieved; the breakdown of the $\chi^2$
contribution from various data sets is given in Table~\ref{Table}.
In the following, we use the data on the reaction $\gamma p \rightarrow K^+\Lambda$ only.

\subsection{Energy-independent PWA with $L=0,1$ multipoles}

It is natural to assume that in the energy region not far above the
threshold only multipoles of low spin play a role. The energy
dependent PWA \cite{Anisovich:2012ct} supports this assumption: in
the region up to 2000 MeV there are four large multipoles, $E_{0+}$,
$E_{1+}$, $M_{1+}$ and $M_{1-}$ which are 5 to 10 times larger than
multipoles with $L=2$.

\begin{figure}[pt]
\begin{center}\vspace{6mm}
\begin{tabular}{cc}
\hspace{-2mm}\includegraphics[width=0.22\textwidth]{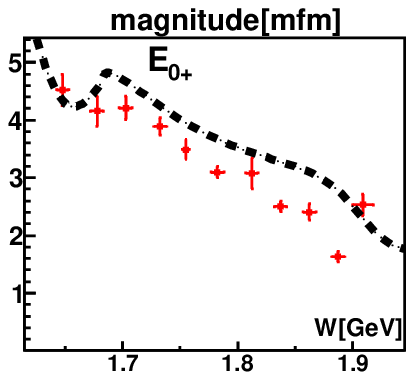}&
\hspace{-2mm}\includegraphics[width=0.22\textwidth]{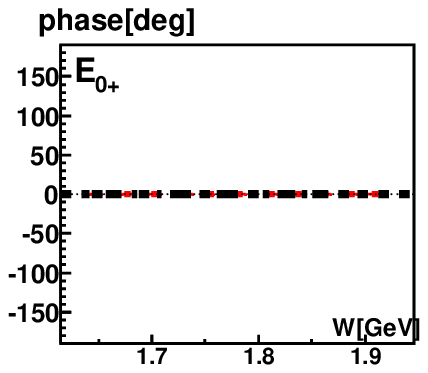}\\
\hspace{-2mm}\includegraphics[width=0.22\textwidth]{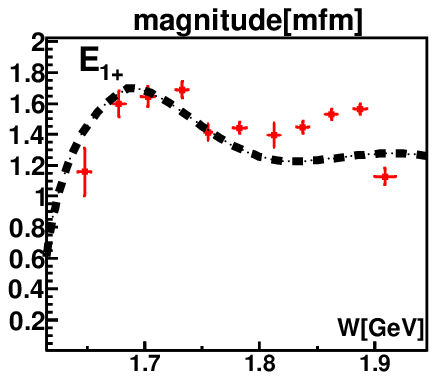}&
\hspace{-2mm}\includegraphics[width=0.22\textwidth]{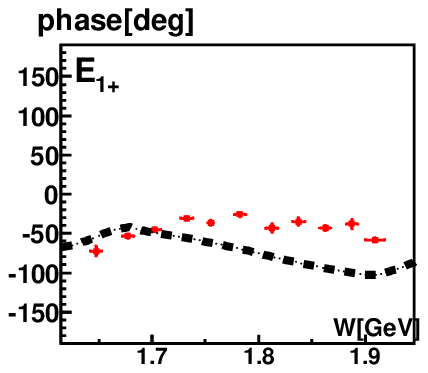}\\
\hspace{-2mm}\includegraphics[width=0.22\textwidth]{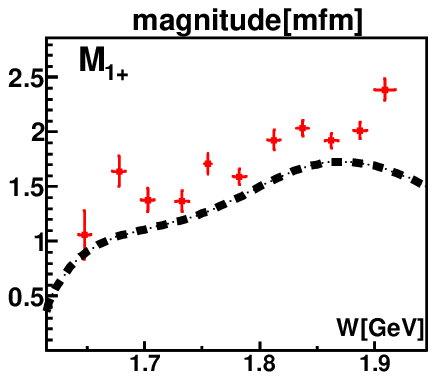}&
\hspace{-2mm}\includegraphics[width=0.22\textwidth]{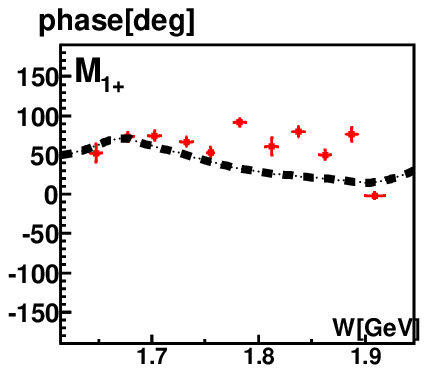}\\
\hspace{-2mm}\includegraphics[width=0.22\textwidth]{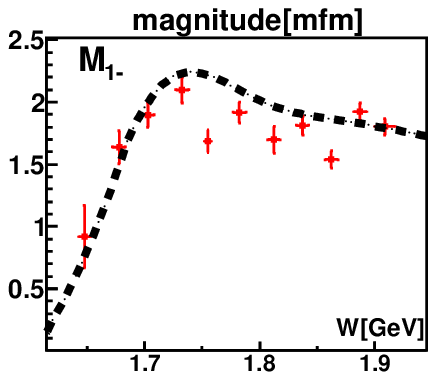}&
\hspace{-2mm}\includegraphics[width=0.22\textwidth]{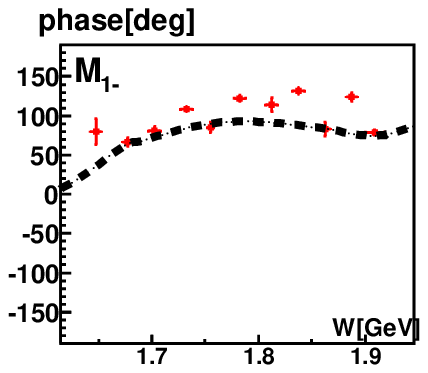}\\
\end{tabular}
\end{center}
\caption{\label{SPwaves}Decomposition of the $\gamma p\to 
K^+\Lambda$ amplitude with $S$ and $P$ multipoles. The general phase is not
defined so we choose $\phi(E^+_0) =0$. The dashed line is the energy
dependent solution BnGa2013. }\vspace{-4mm}
\end{figure}

The multipole decomposition is shown in Fig.~\ref{SPwaves}. In the
fit, it is assumed that only the four multipoles shown in the figure
contribute to the reaction $\gamma p \rightarrow K^+\Lambda$, all other contributions are set to
zero. The errors of the multipoles correspond to changes in
description of the data by one unit in of $\chi^2$. Let us note that
the phases of the multipoles in a fit are defined up to one overall
phase. Here, we determine the phases relative to the phase of the
$E_{0+}$ multipole. Hence $\phi(E^+_0) =0$ holds by construction.  The
comparison with the Bonn-Gatchina PWA shows that the energy-dependent
fit is approximately compatible with the single-energy fit, at least
in the region up to 1750\,MeV. We also note that a truncated PWA with
$L=0,1$ multipoles gives a good description of the data up to this
energy. At $W>1800$ MeV the $L=2$ multipoles are important, see
Fig.~\ref{Observables}. The quality of this fit (in terms of $\chi^2$)
is shown in Table~\ref{Table}. While the differential cross sections
are described very reasonably, the fit to polarization observables is
not convincing: in particular the beam asymmetry is poorly reproduced
and several other polarization variables have $\chi^2$ values
exceeding 2. A more detailed view reveals that the predicted beam
asymmetry $\Sigma$ and the data have a different angular dependence;
this difference is rather pronounced in the mass region above
1800\,MeV. Obviously, a fit with only $L=0$ and $L=1$ multipoles is not
sufficient to describe the data over the full mass range.

%\end{figure}
%\begin{figure}[pt]
%\vspace{4mm}{\bf Fig. 2. continued.
%\end{figure}

\subsection{Energy-independent PWA with $L=0,1,2$ multipoles}

The multipoles with $L=2$ significantly improve the fit quality.
The mean $\chi^2$ per data point drops from 1.8 to 0.8 (see Table~\ref{Table}). 
The improvement is particularly large for the GRAAL beam asymmetry where the $\chi^2$ goes down from 6.77 to 0.57. Most observables are now fitted with a $\chi^2$ per data point of less than 1. The number of fit parameters (moduli of 8 amplitudes and 7 phases at 11 energies) is 165. It is likely that
the systematic errors given in the publications are slightly overestimated. The improvement of the fit can also be seen when Fig.~\ref{Observables} is inspected. 

The resulting multipole decomposition is shown in the two left columns
of Fig.~\ref{SPDwaves}. We observe that the multipoles scatter from
bin to bin. Moreover, for some energy bins there are no $C_x$ and
$C_z$ data. The solution is no longer uniquely defined: two different
solutions are found which differ less than $\delta \chi^2
< 1$. Two conclusions follow from these observations: i) at energies
$W>1750$ MeV the $L=2$ multipoles are definitely needed. ii) the lack
of experimental data and the data quality does not allow extraction of
multipoles with the desired precision in a completely free fit.

\begin{table*}[pt] \caption{\label{Table} Quality of the
energy-independent fit: $\chi^2/N_{\rm data}$ and number of data
points (in brackets). }
\renewcommand{\arraystretch}{1.2}
\begin{center}
\begin{tabular}{cccccc}
  \hline\hline
  Data & BnGa & EI PWA & EI PWA   &EI PWA   & BnGa2013\\
       & 2013 & $L=0,1$& $L=0,1,2$& penalty & penalty\\
  \hline
  $d\sigma/d\Omega$ (CLAS+GRAAL) & 1.85 (316)   & 1.15 & 0.81 & 0.82 & 0.85\\
  $d\sigma/d\Omega$ (MAMI)       & 1.55 (510)   & 1.05 & 0.84 & 0.87 & 0.87\\
  $\Sigma$ (GRAAL)               & 2.44 (66)    & 6.77 & 0.57 & 2.08 & 0.81\\
  $P$ (CLAS)                     & 1.2  (184)   & 3.02 & 0.80 & 1.03 & 0.86\\
  $P$ (GRAAL)                    & 0.65 (66)     & 2.49 & 0.68 & 1.27 & 0.64\\
  $T$                            & 1.54 (66) &  2.02    & 0.61 & 1.20 & 0.98\\
  $O_{x'}$                       & 1.73 (66) &  3.29    & 0.42 & 1.53 & 1.24\\
  $O_{z'}$                       & 1.88 (66) &  2.68    & 0.81 & 1.13 & 1.19\\
  $C_{x}$                        & 1.64 (43) &  1.10    & 0.99 & 1.17 & 1.01\\
  $C_{z}$                        & 1.65 (43) &  1.94    & 1.39 & 2.49 & 1.36\\
\hline Mean                      & 1.57     & 1.8      & 0.8 & 0.9  & 0.85\\
  \hline\hline
\end{tabular}
\end{center}\vspace{-6mm}
\renewcommand{\arraystretch}{1.0}
\end{table*}

\subsection{Energy-independent PWA with $L=0,1,2$ multipoles and penalty function }

In a next step we guide the fit with $L=0,1,2$ multipoles so it is not
totally free. We assume that the large multipoles with $L=0,1$ are
reasonably well defined by the fit using $L=0,1$ multipoles only. Thus
we impose a penalty function which sanctions solutions which deviate
strongly from the fit with $L=0,1$ multipoles.  More precisely, we
introduce a penalty function defined as

\be
\chi^2_{pen} = \sum_{\alpha} \frac{(M_\alpha - 
M_\alpha^{0,1})^2}{(\delta  M_\alpha^{0,1})^2} +\sum_{\alpha}
\frac{(E_\alpha-  E_\alpha^{0,1})^2}{(\delta  E_\alpha^{0,1})^2}\,,
\label{penalty_1}
\ee
where $E_\alpha^{0,1}$ and $M_\alpha^{0,1}$ are the electric and
magnetic multipoles from solution with $L=0,1$ multipoles only;
$\delta E_\alpha^{0,1}$, $\delta M_\alpha^{0,1}$ are the multipole
uncertainties.

The quality of the fit to the differential cross sections hardly
changes while most polarization observables are now described
 with lesser accuracy (see Table~\ref{Table}).

\begin{figure*}[pt]
\begin{center}
\begin{tabular}{cccccc}
\hspace{-2mm}\includegraphics[width=0.17\textwidth]{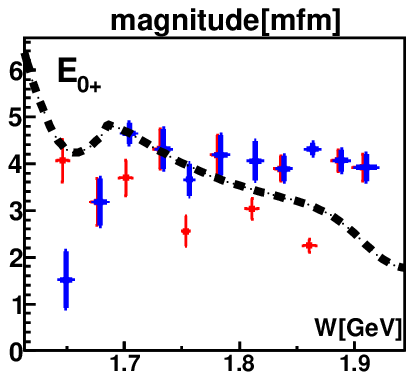}&
\hspace{-5mm}\includegraphics[width=0.17\textwidth]{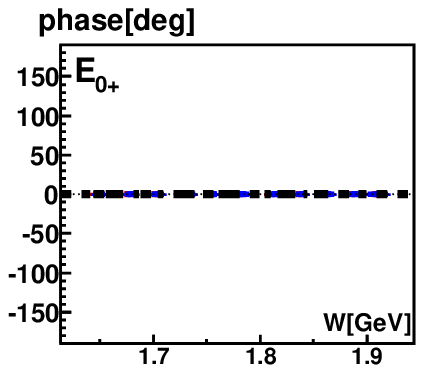}&
\hspace{-5mm}\includegraphics[width=0.17\textwidth]{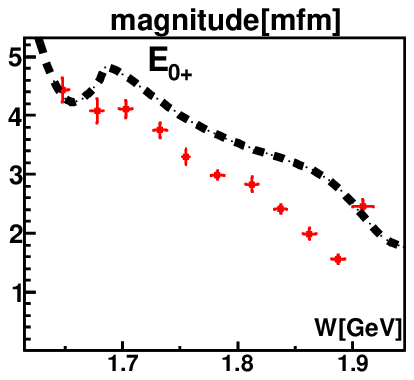}&
\hspace{-5mm}\includegraphics[width=0.17\textwidth]{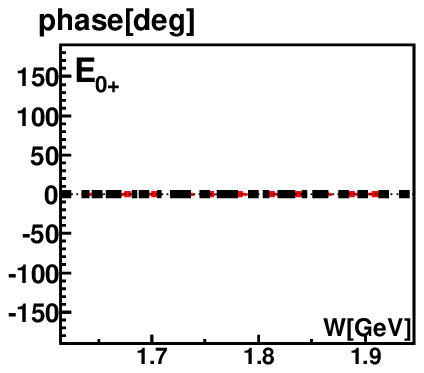}&
\hspace{-5mm}\includegraphics[width=0.17\textwidth]{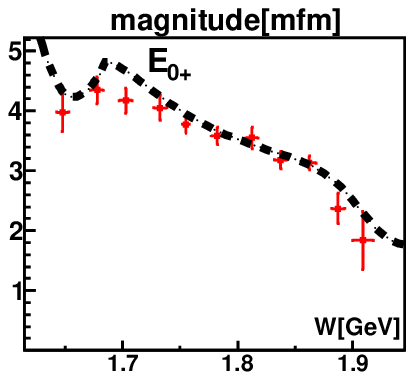}&
\hspace{-5mm}\includegraphics[width=0.17\textwidth]{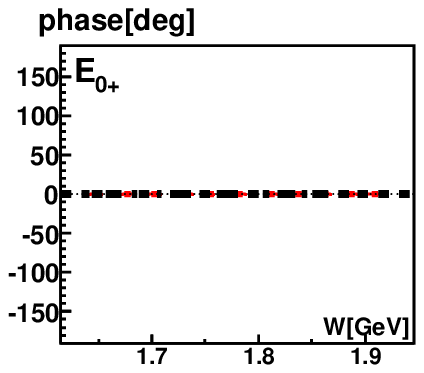}\\
\hspace{-2mm}\includegraphics[width=0.17\textwidth]{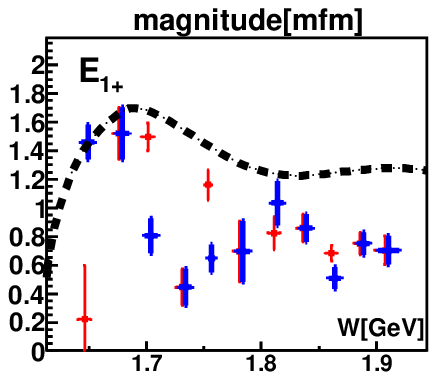}&
\hspace{-5mm}\includegraphics[width=0.17\textwidth]{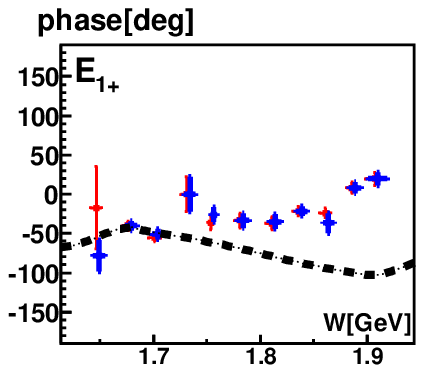}&
\hspace{-5mm}\includegraphics[width=0.17\textwidth]{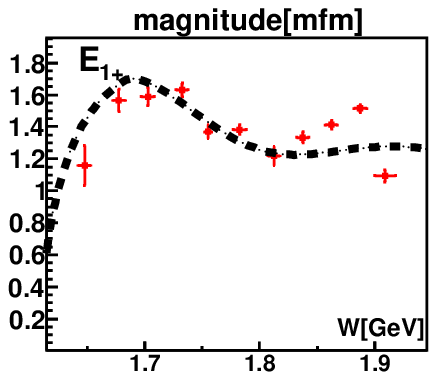}&
\hspace{-5mm}\includegraphics[width=0.17\textwidth]{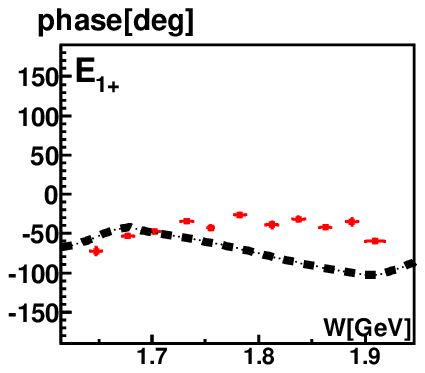}&
\hspace{-5mm}\includegraphics[width=0.17\textwidth]{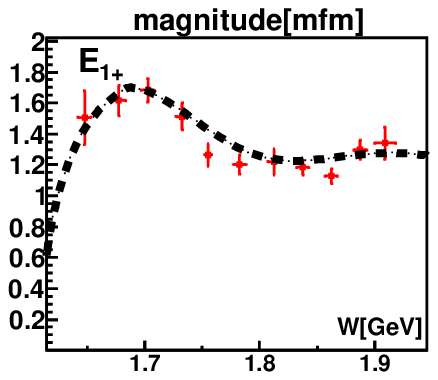}&
\hspace{-5mm}\includegraphics[width=0.17\textwidth]{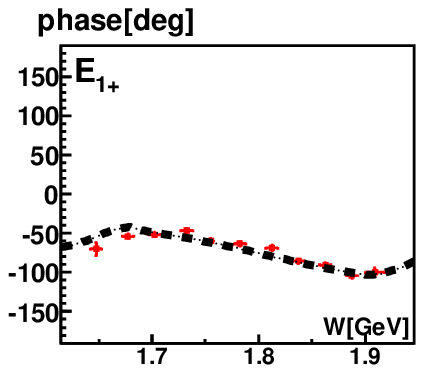}\\
\hspace{-2mm}\includegraphics[width=0.17\textwidth]{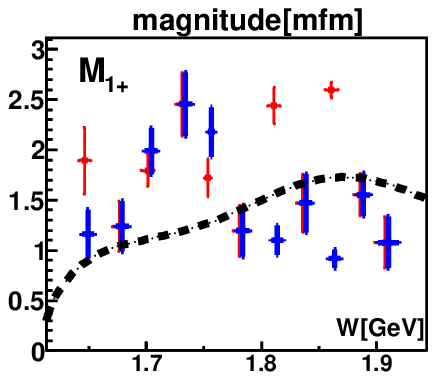}&
\hspace{-5mm}\includegraphics[width=0.17\textwidth]{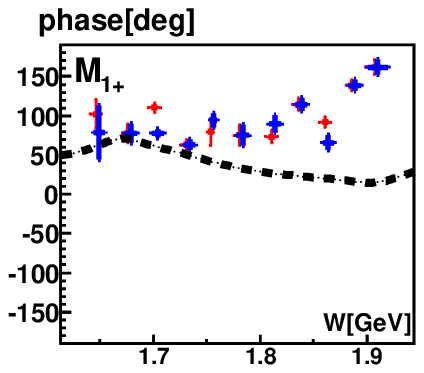}&
\hspace{-5mm}\includegraphics[width=0.17\textwidth]{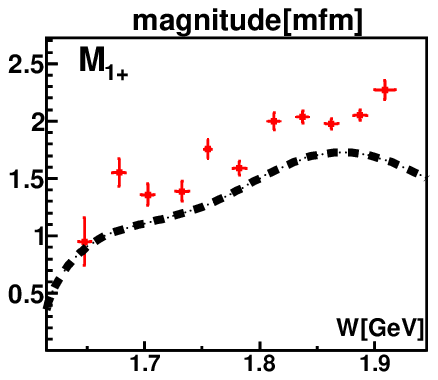}&
\hspace{-5mm}\includegraphics[width=0.17\textwidth]{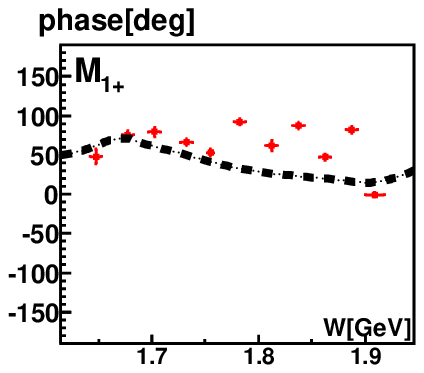}&
\hspace{-5mm}\includegraphics[width=0.17\textwidth]{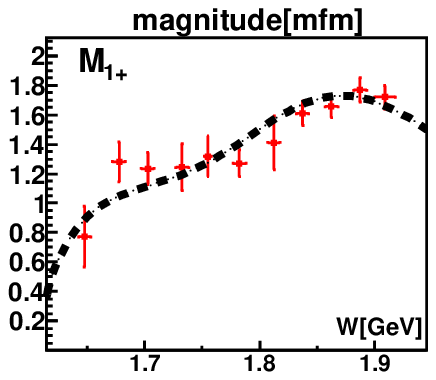}&
\hspace{-5mm}\includegraphics[width=0.17\textwidth]{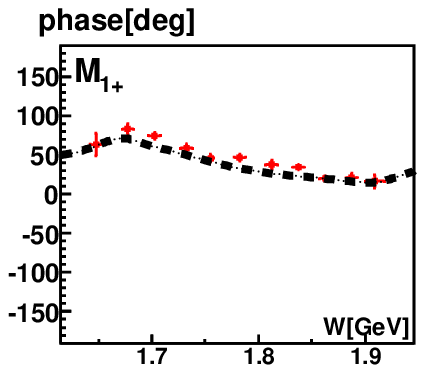}\\
\hspace{-2mm}\includegraphics[width=0.17\textwidth]{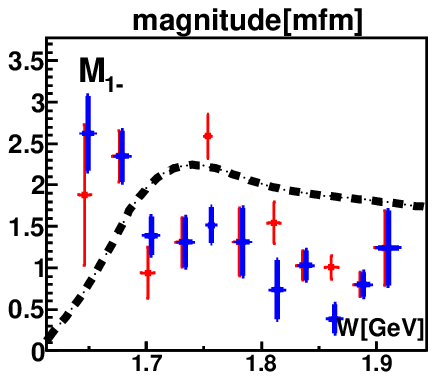}&
\hspace{-5mm}\includegraphics[width=0.17\textwidth]{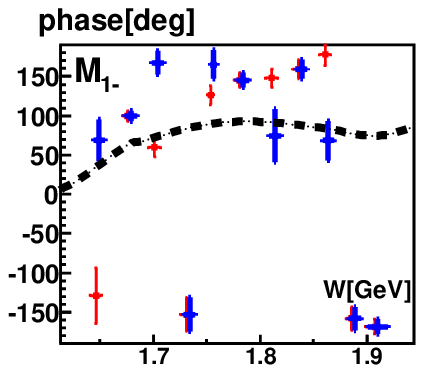}&
\hspace{-5mm}\includegraphics[width=0.17\textwidth]{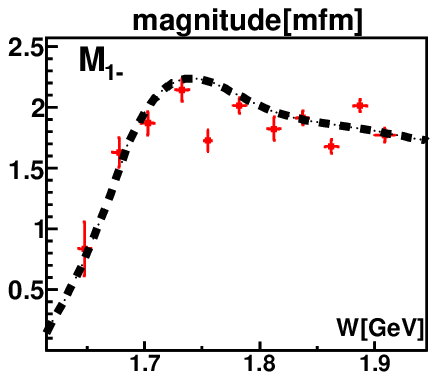}&
\hspace{-5mm}\includegraphics[width=0.17\textwidth]{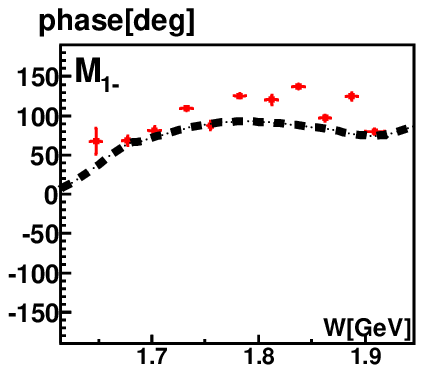}&
\hspace{-5mm}\includegraphics[width=0.17\textwidth]{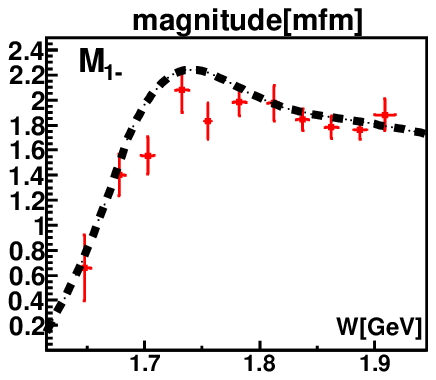}&
\hspace{-5mm}\includegraphics[width=0.17\textwidth]{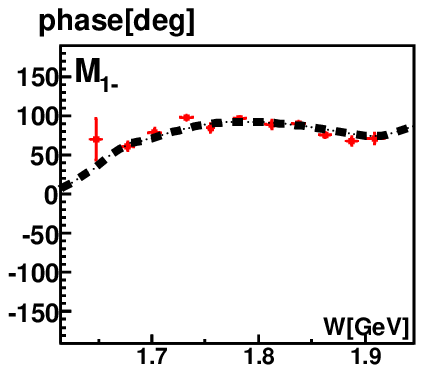}\\
\hspace{-2mm}\includegraphics[width=0.17\textwidth]{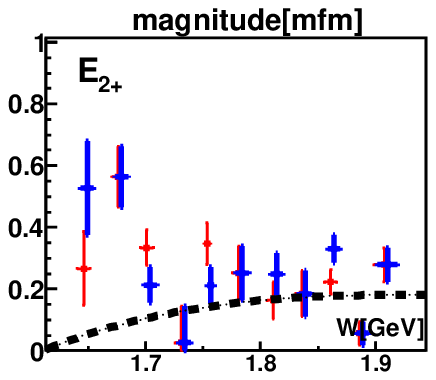}&
\hspace{-5mm}\includegraphics[width=0.17\textwidth]{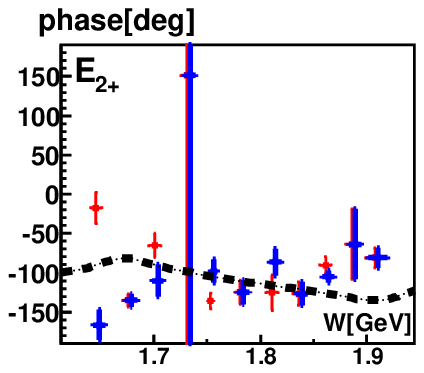}&
\hspace{-5mm}\includegraphics[width=0.17\textwidth]{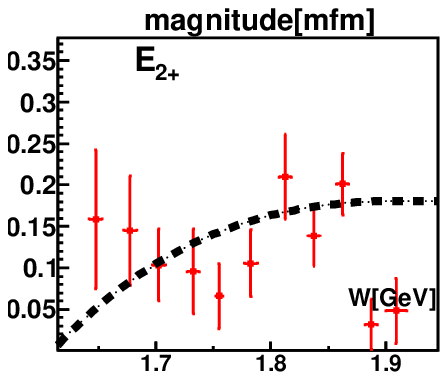}&
\hspace{-5mm}\includegraphics[width=0.17\textwidth]{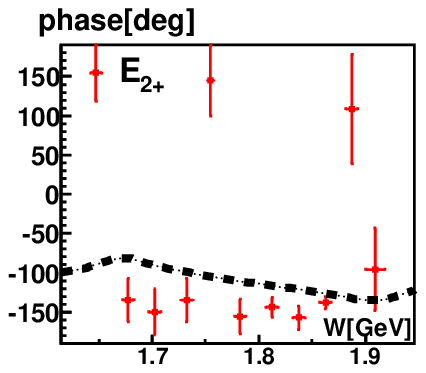}&
\hspace{-5mm}\includegraphics[width=0.17\textwidth]{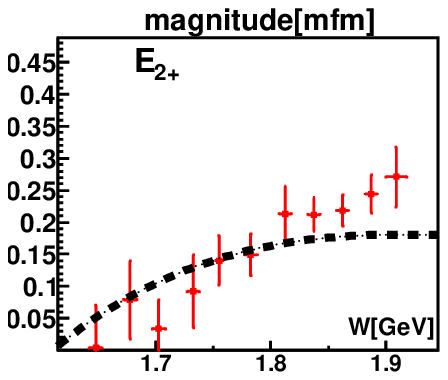}&
\hspace{-5mm}\includegraphics[width=0.17\textwidth]{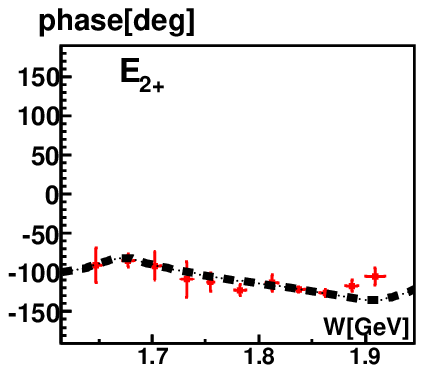}\\
\hspace{-2mm}\includegraphics[width=0.17\textwidth]{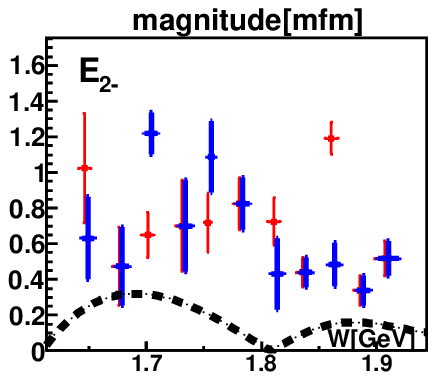}&
\hspace{-5mm}\includegraphics[width=0.17\textwidth]{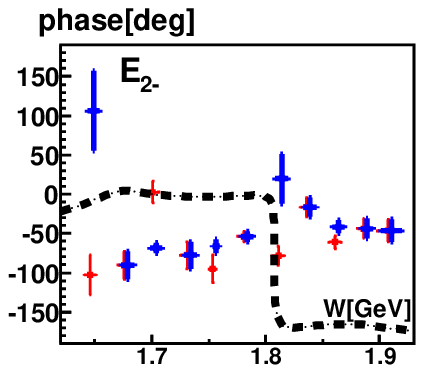}&
\hspace{-5mm}\includegraphics[width=0.17\textwidth]{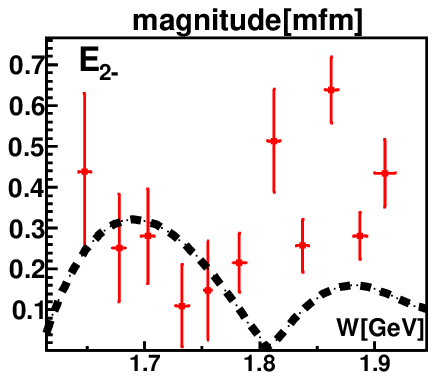}&
\hspace{-5mm}\includegraphics[width=0.17\textwidth]{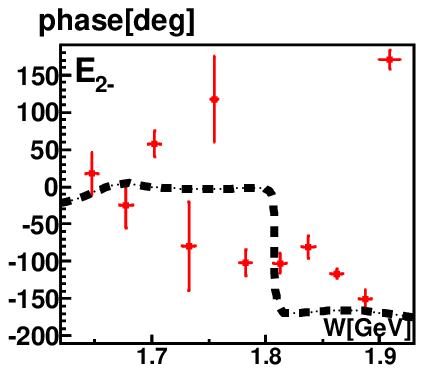}&
\hspace{-5mm}\includegraphics[width=0.17\textwidth]{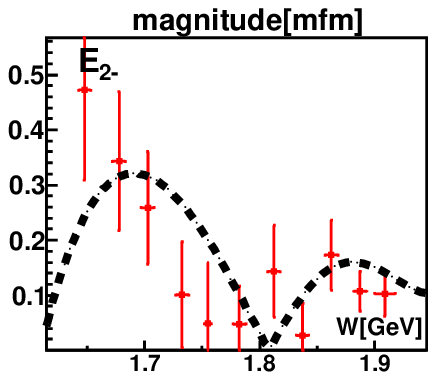}&
\hspace{-5mm}\includegraphics[width=0.17\textwidth]{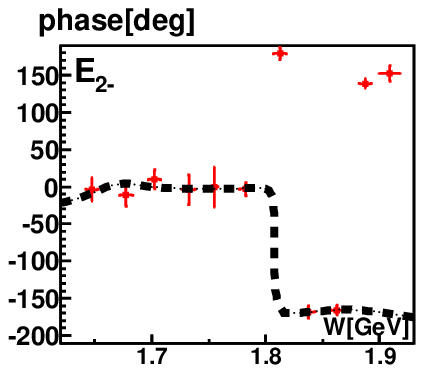}\\
\hspace{-2mm}\includegraphics[width=0.17\textwidth]{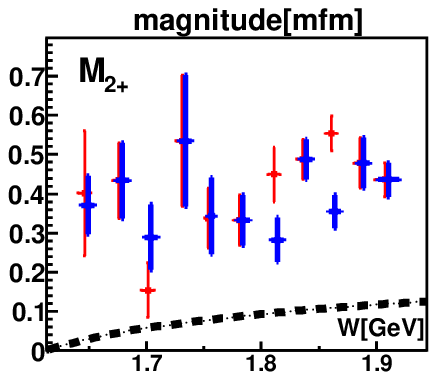}&
\hspace{-5mm}\includegraphics[width=0.17\textwidth]{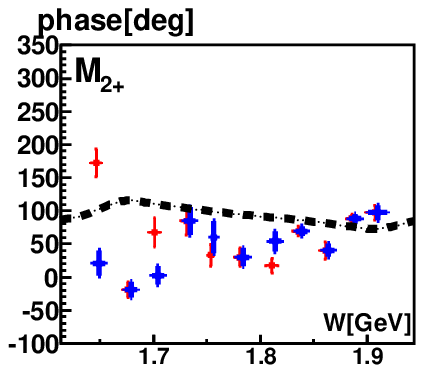}&
\hspace{-5mm}\includegraphics[width=0.17\textwidth]{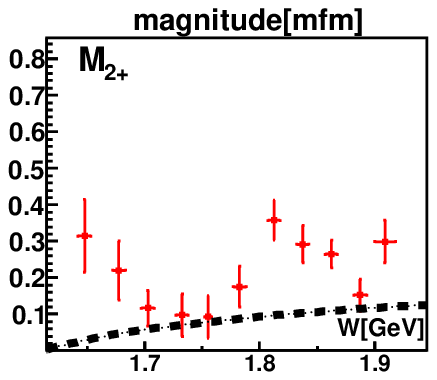}&
\hspace{-5mm}\includegraphics[width=0.17\textwidth]{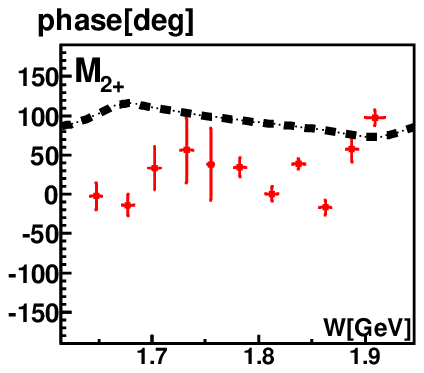}&
\hspace{-5mm}\includegraphics[width=0.17\textwidth]{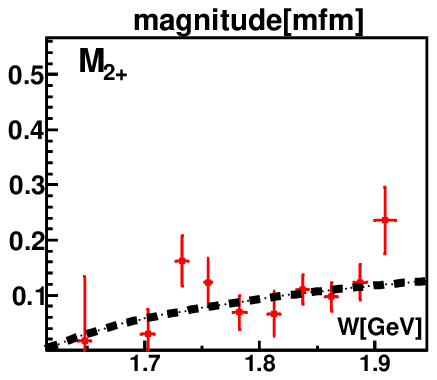}&
\hspace{-5mm}\includegraphics[width=0.17\textwidth]{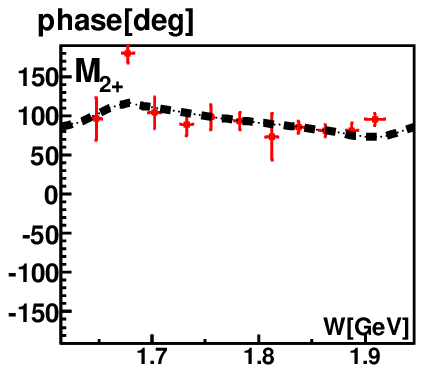}\\
\hspace{-2mm}\includegraphics[width=0.17\textwidth]{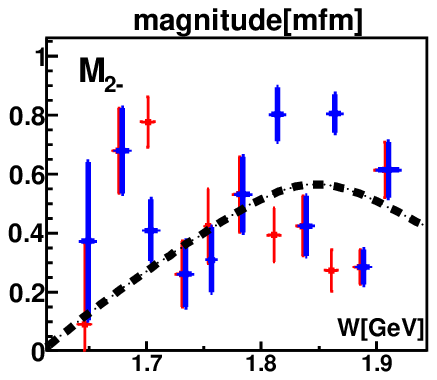}&
\hspace{-5mm}\includegraphics[width=0.17\textwidth]{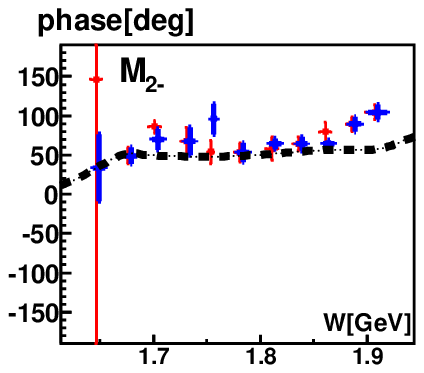}&
\hspace{-5mm}\includegraphics[width=0.17\textwidth]{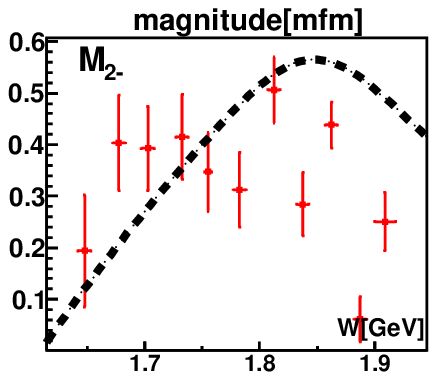}&
\hspace{-5mm}\includegraphics[width=0.17\textwidth]{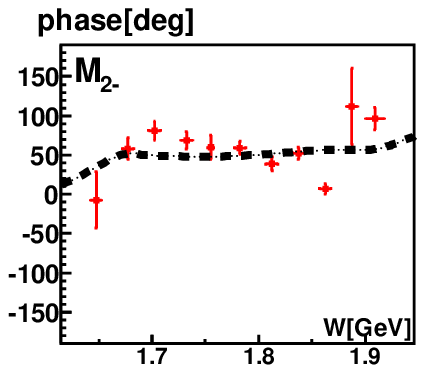}&
\hspace{-5mm}\includegraphics[width=0.17\textwidth]{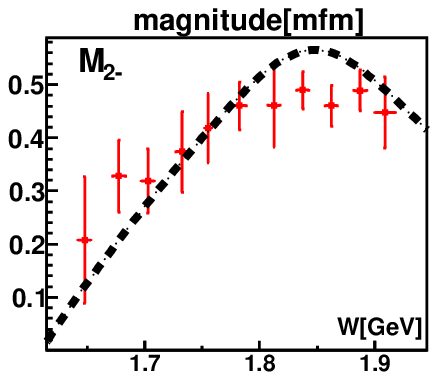}&
\hspace{-5mm}\includegraphics[width=0.17\textwidth]{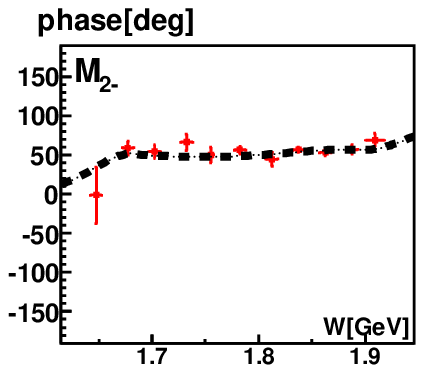}
\end{tabular}
\end{center}
\caption{\label{SPDwaves}Decomposition of the $\gamma p\to K^+\Lambda
$ amplitude with $S$, $P$, and $D$ multipoles. In the low-energy region, two solutions (red
and blue) exist which give identical fits to the data. The dashed line is the energy-dependent solution
BnGa2013. The left two columns represent a free fit. In the two
center columns, the penalty function (Eq.~\ref{penalty_1}) is used. For
the two columns on the right, (Eq.~\ref{penalty_2}) is
used.}\vspace{-14mm}
\end{figure*}

The resulting multipoles are shown in the two center columns of
Fig.~\ref{SPDwaves}. There is now a unique solution but the solution
still scatters significantly for the small multipoles, and the
$E_{0^+}$ and $M_{1^+}$ waves deviate significantly from the the
energy-dependent fit. It has to be stressed that so far, the
solution has no bias at all; the solution is constructed from the
experimental data without any input from the energy-dependent
solution. Even though there are discrepancies between the
energy-dependent and independent solution in detail, the overall
agreement is very satisfactory. In particular there is no hint that
an additional narrow resonance may be hidden or that too many
resonances have been used to fit the data.

\subsection{Constrained energy-independent PWA as a test of the energy
dependent solution}

The energy-independent solution can be used as a test of the
energy-dependent solution BnGa2013 PWA \cite{Anisovich:2012ct}. The
goals are to to check the stability of energy-dependent multipoles
$L=0,1,2$ and to search for any missing structures. Thus the penalty
function, Eq.~\ref{penalty_2}, is included in the fit to control the deviation of
$L=0,1,2$ multipoles from the BnGa2013 solution:
\be
\hspace{-3mm}\chi^2_{pen} = \sum_{\alpha} \frac{(M_\alpha - 
M_\alpha^{0,1,2})^2}{(\delta M_\alpha^{0,1,2})^2} +\sum_{\alpha}
\frac{(E_\alpha-  E_\alpha^{0,1,2})^2}{(\delta 
  E_\alpha^{0,1,2})^2}\,,
\label{penalty_2}
\ee
where
$E_\alpha^{0,1,2}$ and $M_\alpha^{0,1,2}$ are the
multipoles for BnGa2013 $L=0,1,2$ solution and $\delta{
E_\alpha^{0,1,2}}$, $\delta {M_\alpha^{0,1,2}}$ are the multipole
uncertainties for the fit without penalty. In this approach,
multipoles with $L\ge 3$ are fixed by the energy-dependent solution.
The error in the multipoles from the BnGa2013 energy-dependent solution
is not included in the definition of the penalty function. The result
of the fit is shown in the two columns on the right in
Fig.~\ref{SPDwaves}.

The fit is only marginally worse than the unconstrained fit.
This proves the quality of the energy-dependent fit.

\section{Conclusion}
\label{sec:conclusions}

We have performed an energy-independent partial-wave analysis for the
reaction $\gamma p\to K^+ \Lambda$ in the region up to an
invariant mass $W=1918$~MeV. Although not yet complete, a data set of
differential cross section values and polarization observables was
available that allowed an energy-independent extraction of the
dominant electromagnetic multipoles that underlie the production
process. The analysis requires multipoles up to $L=2$, and there is no
evidence that the fit requires multipoles with $L\ge 3$.

At present the available data allow for the extraction of multipoles
$E_{0+}$, $E_{1+}$, $M_{1+}$ and $M_{1-}$ only, without using further
constraints. They
are compatible with multipoles obtained in the energy-dependent fit.
Multipoles with $L=2$ could not be extracted unambiguously without
imposing further, albeit rather mild constraints.

The multipoles from the energy-dependent PWA BnGa2013 were checked for
stability in the single-energy fit constrained by BnGa2013 solution.
The resulting multipoles are very close to the original
energy-dependent solution. There is no evidence for any additional
structures which may have escaped in the energy-dependent fit.

These results demonstrate that using cross section and polarization observables for the photoproduction of pseudoscalar mesons can be successfully employed in
energy-independent PWA without additional
constraints, and that the complex multipoles underlying the
production process can be determined with good accuracy.
It is also demonstrated that the multipoles determined in this manner
are consistent with those determined in more strongly constrained
energy-dependent PWA fits.

These results mark an essential step in the ongoing development of
sound procedures in the search for yet-to-be discovered
excited states of the nucleon. Using data from major single
production channels only, the method enables an independent verification
of discovery claims of new excited states in complex and highly
constrained coupled-channel analyses.


\begin{thebibliography}{99}
\bibitem{PDG:2012}
  J.~Beringer et al. (Particle Data Group),
  Phys.~Rev.~D {\bf 86}, 010001 (2012)

\bibitem{Capstick:1993piN}
  S.~Capstick and W.~Roberts,
  %''$N\pi$ decays of baryons in a relativised model,''
  Phys.~Rev.~D {\bf 47}, 1994-2010 (1993).

\bibitem{Capstick:1998D}
  S.~Capstick and W.~Roberts,
  %''New baryons in the $\Delta \eta$ and $\Delta \omega$ channels'',
  Phys.~Rev.~D {\bf 57}, 4301 (1998).
  %doi = {10.1103/PhysRevD.57.4301},
  %url = {http://link.aps.org/doi/10.1103/PhysRevD.57.4301}

\bibitem{Capstick:1998KY}
  S.~Capstick and W.~Roberts,
  %''Strange decays of nonstrange baryons'',
  Phys.~Rev.~D {\bf 58}, 074011 (1998).

%\cite{Chiang:1996em}
\bibitem{Chiang:1996em}
  W.~-T.~Chiang and F.~Tabakin,
  %``Completeness rules for spin observables in pseudoscalar meson photoproduction,''
  Phys.\ Rev.\ C {\bf 55}, 2054 (1997).

\bibitem{Omelaenko:1981}
A.S.~Omelaenko, Sov.J.Nucl.Phys. {\bf 34}, 406 (1981).

\bibitem{Wunderlich:2013iga} 
  Y.~Wunderlich, R.~Beck and L.~Tiator,
  ``Ambiguities of the single spin observables in a truncated partial wave expansion for photoproduction of pseudoscalar mesons,''
  arXiv:1312.0245 [nucl-th].

\bibitem{Hartmann:2014}
Jan Hartmann {\it et al.} [CBELSA/TAPS collaboration],  
`` The $N(1520)$ helicity amplitude from an energy-independent multipole
analysis based on new polarization data'', 
submitted to Phys. Rev. Lett. (2014).

\bibitem{Mart:1999}
  T.~Mart and C.~Bennhold,
  %''Evidence for a missing nucleon resonance in kaon photoproduction'',
  Phys.~Rev.~C {\bf 61}, 012201 (1999).

\bibitem{Glander:2004}
  K. H. Glander {\it et al.}  [SAPHIR Collaboration] ,
  Eur.~Phys.~J.~A {\bf19}, 251 (2004).

\bibitem{Guidal:2003}
  M.~Guidal, J.-M.~Laget, and M.~Vanderhaeghen,
  %''Exclusive electromagnetic production of strangeness on the nucleon: review of recent data in a Regge approach'',
  Phys.~Rev.~C {\bf 68}, 058201 (2003).

\bibitem{Janssen:2001}
  S.~Janssen, J.~Ryckebusch, D.~Debruyne, and T.~Van Cauteren,
  %''Kaon photoproduction: Back-ground contributions, form factors, and missing resonances'',
  Phys.~Rev.~C {\bf 65}, 015201 (2001).

\bibitem{Corthals:2006}
  T.~Corthals, J.~Ryckebusch, and T.~Van Cauteren,
  %''Forward-angle $K^+\Lambda$ photoproduction in a Regge-plus-resonance approach'',
  Phys.~Rev.~C {\bf 73}, 045207 (2006).

%\cite{Schumacher:2010qx}
\bibitem{Schumacher:2010qx}
  R.~A.~Schumacher and M.~M.~Sargsian,
  %``Scaling and Resonances in Elementary K^+ Lambda Photoproduction,''
  Phys.\ Rev.\ C {\bf 83}, 025207 (2011).
  %%CITATION = ARXIV:1012.2126;%%

\bibitem{Bradford:2005pt}
  R.~Bradford {\it et al.}  [CLAS Collaboration],
  %``Differential cross sections for gamma + p --> K+ + Y for $\Lambda$ and  $\Sigma^0$
  %hyperons,''
  Phys.\ Rev.\  C {\bf 73}, 035202 (2006).
%[arXiv:nucl-ex/0509033].

\bibitem{McNabb:2003nf}
  J.~W.~C.~McNabb {\it et al.}  [CLAS Collaboration],
  %``Hyperon photoproduction in the nucleon resonance region,''
  Phys.\ Rev.\ C {\bf 69}, 042201 (2004).
  %%CITATION = NUCL-EX/0305028;%%
  %142 citations counted in INSPIRE as of 04 Apr 2014

%\bibitem{Julia-Diaz:2005}
%  B.~Juli\'{a}́-D\'{i}az, B.~Saghai, F.~Tabakin, W.-T.~Chiang, T.-S.~H.~Lee, and Z.~Li,
%  %''Dynamical coupled-channel analysis of $K^+\Lambda$ photoproduction''
%  Nucl.~Phys.~A {\bf 755}, 463-466 (2005).

\bibitem{Mart:2006dk}
  T.~Mart and A.~Sulaksono, and references therein,
  %``Kaon photoproduction in a multipole approach,''
  Phys.\ Rev.\ C {\bf 74}, 055203 (2006).

\bibitem{Sarantsev:2005}
  A.V.~Sarantsev, V.A.~Nikonov, A.V.~Anisovich, E.~Klempt, U.~Thoma,
  %Decays of baryon resonances into $\Lambda K^+$, $\Sigma K^+$, and $\Sigma^+K^0$'',
  Eur.~Phys.~J.~A {\bf 25}, 3 (2005).

\bibitem{Anisovich:2007cxcz}
  A.V.~Anisovich, V.~Kleber, E.~Klempt, V.A.~Nikonov, A.V.~Sarantsev, U.~Thoma,
  %''Baryon resonances and polarization transfer in hyperon photoproduction'',
  Eur.~Phys.~J.~A {\bf 34}, 243 (2007).

  \bibitem{Bradford:2006ba}
  R.~Bradford {\it et al.}   [CLAS Collaboration],
  %``First measurement of beam-recoil observables C(x) and C(z) in hyperon
  %photoproduction,''
  Phys.\ Rev.\  C {\bf 75}, 035205 (2007).

%\cite{McCracken:2009ra}
\bibitem{McCracken:2009ra}
  M.~E.~McCracken {\it et al.}  [CLAS Collaboration],
  %``Differential cross section and recoil polarization measurements for the
  %gamma p to K+ Lambda reaction using CLAS at Jefferson Lab,''
  Phys.\ Rev.\  C {\bf 81}, 025201 (2010).
  %%CITATION = ARXIV:0912.4274;%%

%\cite{Anisovich:2011fc}
\bibitem{Anisovich:2011fc}
  A.~V.~Anisovich, R.~Beck, E.~Klempt, V.~A.~Nikonov, A.~V.~Sarantsev and U.~Thoma,
  %``Properties of baryon resonances from a multichannel partial wave analysis,''
  Eur.\ Phys.\ J.\ A {\bf 48}, 15 (2012).

%\cite{Anisovich:2012ct}
\bibitem{Anisovich:2012ct}
  A.~V.~Anisovich, R.~Beck, E.~Klempt, V.~A.~Nikonov, A.~V.~Sarantsev and U.~Thoma,
  %``Pion- and photo-induced transition amplitudes to $\Lambda K$, $\Sigma K$, and $N\eta$,''
  Eur.\ Phys.\ J.\ A {\bf 48}, 88 (2012).
  %%CITATION = ARXIV:1205.2255;%%

\bibitem{Anisovich:2013vpa}
  A.~V.~Anisovich, E.~Klempt, V.~A.~Nikonov, A.~V.~Sarantsev and U.~Thoma,
  %``Sign ambiguity in the K Sigma channel,''
 Eur.\ Phys.\ J.\ A {\bf 49}, 158 (2013).

 \bibitem{Lleres:2007tx}
  A.~Lleres {\it et al.}  [GRAAL Collaboration],
  %``Polarization observable measurements for gamma p $\to$ K+ Lambda and gamma
  %p $\to$ K+ Sigma0 for energies up to 1.5-GeV,''
  Eur.\ Phys.\ J.\  A {\bf 31}, 79 (2007).
  %%CITATION = EPHJA,A31,79;%%
%%CITATION = PHRVA,C73,035202;%%

\bibitem{Lleres:2008em}
  A.~Lleres {\it et al.}  [GRAAL Collaboration],
  %``Measurement of beam-recoil observables Ox, Oz and target asymmetry for the
  %reaction gamma p -> K Lambda,''
  Eur.\ Phys.\ J.\  A {\bf 39}, 149 (2009).
  %[arXiv:0807.3839 [nucl-ex]].
  %%CITATION = EPHJA,A39,149;%%

%\cite{Jude:2013jzs}
\bibitem{Jude:2013jzs}
  T.~C.~Jude {\it et al.} [Crystal Ball Collaboration],
  ``$K^+\Lambda$ and $K^+\Sigma^0$ photoproduction with fine center-of-mass energy resolution,''
  arXiv:1308.5659 [nucl-ex].
  %%CITATION = ARXIV:1308.5659;%%

%\cite{Drechsel:1998hk}
\bibitem{Drechsel:1998hk}
  D.~Drechsel, O.~Hanstein, S.~S.~Kamalov and L.~Tiator,
  %``A Unitary isobar model for pion photoproduction and electroproduction on the proton up to 1-GeV,''
  Nucl.\ Phys.\ A {\bf 645}, 145 (1999).

%\cite{Fasano:1992es}
\bibitem{Fasano:1992es}
  C.~G.~Fasano, F.~Tabakin and B.~Saghai,
  %``Spin observables at threshold for meson photoproduction,''
  Phys.\ Rev.\ C {\bf 46}, 2430 (1992).
  %%CITATION = PHRVA,C46,2430;%%

\end{thebibliography}
\end{document}